\documentclass[nofootinbib,aps,pra,superscriptaddress,onecolumn, titlepage=false]{revtex4-2}

\usepackage{graphicx}
\usepackage{amsmath}
\usepackage{mathrsfs}
\usepackage{dsfont}
\usepackage{amssymb}
\usepackage{subfigure}
\usepackage{color}
\usepackage{blkarray}
\usepackage{hyperref}
\usepackage{verbatim}
\usepackage{amsthm}
\usepackage{enumerate}
\usepackage{bbm}
\usepackage{hyperref}
\usepackage{braket}
\usepackage{booktabs}
\usepackage{mathtools}
\usepackage{titlesec}
\usepackage{array,booktabs}

\newcommand{\virg}[1]{``#1''}
\newcommand{\eq}[1]{Eq.~\eqref{#1}}

\newcommand*{\h}{\mathcal{H}}
\newcommand*{\Tr}{\mathrm{Tr}}
\newcommand*{\diff}{\text{d}}

\newcommand*{\coloneqq}{\mathrel{\vcenter{\baselineskip0.5ex \lineskiplimit0pt \hbox{\scriptsize.}\hbox{\scriptsize.}}} =}
\newcommand*{\coloneqqrev}{=\mathrel{\vcenter{\baselineskip0.5ex \lineskiplimit0pt \hbox{\scriptsize.}\hbox{\scriptsize.}}} }

\begin{document}
		\title{\Large Holographic maps from quantum gravity states as tensor networks}
		\author{Eugenia Colafranceschi}
	\email{eugenia.colafranceschi@nottingham.ac.uk}
	\affiliation{School of Mathematical Sciences and Centre for the Mathematics and Theoretical Physics of Quantum Non-Equilibrium Systems, University of Nottingham, University Park Campus, Nottingham NG7 2RD, United Kingdom}
	
	\author{Goffredo Chirco}
	\email{goffredo.chirco@na.infn.it}
	\affiliation{Istituto Nazionale di Fisica Nucleare - Sezione di Napoli, Complesso Universitario di Monte S. Angelo ed. 6, Via Cintia, 80126 Napoli, Italia}

	\author{Daniele Oriti}
	\email{daniele.oriti@physik.lmu.de}
	\affiliation{Arnold Sommerfeld Center for Theoretical Physics, \\ Ludwig-Maximilians-Universit\"at München \\ Theresienstrasse 37, 80333 M\"unchen, Germany}

\begin{abstract}
We define bulk/boundary maps corresponding to quantum gravity states in the tensorial group field theory formalism, for quantum geometric models sharing the same type of quantum states of loop quantum gravity. The maps are defined in terms of a partition of the quantum geometric data associated to an open graph into bulk and boundary ones, in the spin representation. We determine the general condition on the entanglement structure of the state that makes the bulk/boundary map isometric (a necessary condition for holographic behaviour), and we analyse different types of quantum states, identifying those that define isometric bulk-to-boundary maps.
\end{abstract}
	\maketitle

	\section*{Introduction}		\addcontentsline{toc}{section}{Introduction}

The idea of holography, broadly understood as the physical situation in which the entire properties of a system occupying a certain region of space are in fact fully encoded in the boundary of the same region, has become more and more central in theoretical physics, over the last 25 years. The first example of holographic behaviour was provided by black holes~\cite{Bekenstein:1975tw, tHooft:1984kcu, Susskind:1994vu}, with the area law for their entropy, leading to the suggestion that their microscopic degrees of freedom actually reside on the horizon, rather than being distributed across their interior, or at least that they can be mapped there\footnote{The nature of microscopic black hole degrees of freedom and of their entropy is still a hotly debated issue, though.}. Being black holes peculiar configurations of the gravitational field and of spacetime geometry, holography was immediately suggested to be a key element in understanding spacetime geometry and gravitational physics at a more fundamental level, in particular for what concerns the causal structure of the world~\cite{tHooft:1993dmi,Bousso:2002ju}, and probably an important tool to unravel the mysteries of quantum gravity. 

In fact, the next context in which the holographic idea was instantiated was a quantum gravity-related one, i.e. the AdS/CFT correspondence with the conjectured duality between a gravitational theory (e.g. string theory) in a bulk spacetime with asymptotically anti-de Sitter boundary conditions and a conformal field theory (e.g. supersymmetric Yang-Mills) on its flat boundary~\cite{Maldacena:1997re}. This is also the context in which most current work on holographic behaviour is carried out, with the bulk/boundary correspondence being implemented in a large variety of systems and studied from very diverse angles~\cite{Pastawski:2015qua,Almheiri:2014lwa,Bao:2015uaa,Yang:2015uoa,Bhattacharyya:2016hbx,Miyaji:2016mxg,Bao:2019fpq,Bao:2018pvs}. However, on the one hand some features of the AdS/CFT correspondence have been suggested to follow directly from general quantum gravity considerations, when AdS boundary conditions apply~\cite{Raju:2019qjq}; on the other, similar bulk/boundary correspondences have been proposed in the case of de Sitter or flat boundary conditions, and holographic behaviour has been studied in quantum gravity formalisms not directly related to string theory~\cite{Dittrich:2018xuk,Dittrich:2017hnl}. It is therefore clear that holography may play a central role in quantum gravity more generally, and viceversa more general quantum gravity considerations are needed to understand origin and meaning of holographic behaviour. 

In fact, we have by now abundant evidence that holographic behaviour can be traced down to the structure of quantum correlations of the fundamental degrees of freedom living on bulk and boundary, i.e. entanglement. The evidence comes again from various corners. Several quantum many-body systems manifest holographic properties in their ground states and the area law shows up, for example, when computing their entanglement entropy~\cite{Eisert:2008ur}. Holographic behaviour, indeed, seems to emerge in correspondence with the maximal entanglement measures. Correspondingly, quantum information techniques for controlling entanglement in quantum many-body systems have become central in the study of holographic behaviour in (quantum) gravitational systems as well~\cite{Pastawski:2015qua,Almheiri:2014lwa,Bao:2015uaa,Yang:2015uoa,Bhattacharyya:2016hbx,Miyaji:2016mxg,Bao:2019fpq,Bao:2018pvs}. To close the circle, with holography being suggested to hold the key to unravel the nature of geometry and spacetime itself, and entanglement correlations being recognised as the origin of holographic behaviour, it is then entanglement that is now indicated as the material that threads spacetime and geometry into existence~\cite{Bianchi:2012ev}. A number of measures of entanglement, in fact, have been shown to admit a geometric interpretation, both in the AdS/CFT context~\cite{Swingle:2009bg,Swingle:2012wq,Vidal:2007hda,Ryu:2006bv,Ryu:2006ef} and beyond, and quantum gravity formalisms in which spacetime is emergent from non-spatiotemporal fundamental quantum entities~\cite{Oriti:2018dsg} should then focus on their entanglement properties to reconstruct geometry out of them.

This is the background that motivates a number of recent works focusing on the definition of bulk/boundary maps at the level of the quantum states of bulk and boundary theories (including dynamical considerations or remaining at the kinematical level), and on the identifications of the conditions under which these maps become holographic. In particular, an important property that is necessary for holographic behaviour in quantum systems is the isometry of the bulk/boundary map, since this ensures that matrix elements (and expectation values) of quantum observables are assigned the same value by bulk and boundary theories, which is necessary condition for a proper duality. Also, when this isometry is not to be exact, for example because it results from some symmetry of the bulk/boundary system, then it becomes an interesting issue to determine the conditions under which it can be approximately true, since these may also identify the regime in which one has emergent holographic dualities at an approximate level \cite{deHaro:2015pia}. 

We ask these questions in the context of quantum gravity states, and in particular within the tensorial group field theory formalism \cite{Oriti:2011jm,Krajewski:2012aw,Carrozza:2016vsq,Oriti:2014uga,Rivasseau:2012yp, Rivasseau:2016zco, Rivasseau:2016wvy, Delporte:2018iyf}; more specifically, we work with a subclass of models endowed with distinctive quantum geometric data, usually called simply group field theories \cite{Oriti:2011jm,Krajewski:2012aw, Oriti:2013aqa, Oriti:2014uga}. The Hilbert space of quantum states of such models is a Fock space, with elementary quanta described as spin network vertices, i.e. nodes with attached open links, which are labelled by irreducible representation of a (Lie) group (or by a group element), with the nodes labelled instead by intertwiners for the same group representations. The groups chosen in most quantum gravity applications are the Lorentz group (or its euclidean counterpart) and its rotation subgroup $SU(2)$. We will use $SU(2)$ in the following, and restrict ourselves to 4-valent vertices. Equivalently, the same quanta can be understood as quantized simplices labelled by the same group-theoretic data, encoding their quantum geometry (i.e. the value of geometric quantities for the simplices, such as face areas and volume, can be computed as functions of such data). We consider the case in which the GFT quanta are (dual to) 3-simplices, i.e. tetrahedra, as apt to be used in 4d quantum gravity models. Generic quantum states can then be described as (superpositions of) spin networks associated to (possibly very complex) graphs, obtained by appropriately \virg{gluing} the spin network vertices, and labelled by the same type of algebraic data, or equivalently as simplicial complexes obtained by gluing of the elementary quantized simplices. The same kind of quantum states, although organized in a different Hilbert space, is shared by canonical loop quantum gravity \cite{Ashtekar:2021kfp,Thiemann:2007zz}. The differences between the Hilbert spaces of group field theory and loop quantum gravity manifest themselves when considering quantum states associated to different graphs. In this work we only consider quantum states for fixed combinatorial structure, thus our results are valid in both contexts. The quantum dynamics of such states is expressed in terms of superpositions of elementary interaction processes among simplices, producing a simplicial complex of one dimension higher (i.e. a simplicial 4-complex, if the fundamental states are gluings of 3-simplices), with quantum amplitudes obtained by summing over their associated algebraic data. These amplitudes take the form of spin foam amplitudes when expressed in terms of group representations, of lattice gauge theories when expressed in terms of group elements, or of simplicial gravity path integral \cite{Finocchiaro:2018hks}. In the group field theory formalism they arise as Feynman amplitudes in the perturbative expansion of the theory. One main goal becomes then to reconstruct continuum spacetime and geometry and effective gravitational physics out of these discrete and purely algebraic structures, as a concrete example of spacetime emergence from non-spatiotemporal quantum gravity entities~\cite{Oriti:2018dsg, Chirco:2018fns}.

The role of entanglement as the \virg{glue} to be used in building up spacetime and geometry becomes apparent already at this discrete level. The pairing of fundamental simplices (or spin network vertices) to form extended simplicial complexes and, at the dynamical level, their basic interaction processes, are exactly given by entanglement correlations between their individual degrees of freedom. The graphs dual to such complexes are indeed \virg{entanglement graphs}, i.e. a graphical representation of patterns of entanglement correlations, and the quantum states of the theory can in fact be understood as (linear combinations of) generalised (and 2nd quantized) tensor networks, as used in quantum many-body systems to encode quantum correlations in an efficient manner~\cite{Ran_2020,2006quant.ph..8197P, ORUS2014117,Bridgeman_2017}. This correspondence will allow us, in this work, to use random tensor network methods in the computation of measures of entanglement for our quantum gravity states. We will give further details on these points in the next section, referring to~\cite{Colafranceschi:2020ern,PhysRevB.86.195114,Singh_2010} for more. 
Here, we only stress that at this discrete level one also finds explicit examples of the conjectured duality between geometry and entanglement, besides the correspondence between entanglement and topology that the very graph structure represents. The area of triangles in the dual simplicial complex scales with the spin representation associated to them, which also measures the entanglement between the two quantum 3-simplices sharing the given triangle. Also, the volume of each 3-simplex scales with the intertwiner label associated to it, which also measures the entanglement between the four triangles on its boundary. More such relations can be found, including discrete versions of the Ryu-Takanayagi formula \cite{Chirco:2017xjb,Han:2017xwo,Chirco_2018}.

In addition to the entanglement patterns among fundamental \virg{quanta of space}, defining the graph structures, there are in general further quantum correlations between the quantum data living {\it on the graphs themselves} and characterizing the quantum states one can associate to them. The entanglement properties of spin network states have been extensively studied in the quantum gravity literature \cite{Singh_2010,Anza:2017dkd,PhysRevD.92.085045,Livine:2006xk,Anza:2016fix, Feller:2017jqx}, and our present work contributes as well to this line of research, as we are going to see that the bulk/boundary maps one can define using such quantum states depend heavily on the associated algebraic data and on their entanglement properties.   


In this article we present the following results.
First, we consider generic quantum states associated to graphs with both internal and external (open) edges and show how they can be understood as maps between bulk and boundary data. We rely on the following basis-dependent splitting of the graph degrees of freedom:
\begin{itemize}
	\item[-] \textit{boundary}: spins and magnetic numbers of non-contracted edges;
	\item[-] \textit{bulk}: spins of internal links and intertwiners (depending on spins of both internal and boundary edges).
\end{itemize}
We show that a generic spin network state can be understood as defining a map between these two sets: bulk states can be seen as the result of applying a graph-dependent map to certain boundary states, and vice versa. For closely related work, see \cite{Chen:2021vrc}. Entanglement enters this picture at two levels: 1) connectivity of the graph, i.e. the already mentioned entanglement pattern between fundamental spin network vertices; 2) quantum correlations between intertwiners.
While $1)$ underlies the definition of the map, $2)$ determines both the input of the bulk-to-boundary map and (part of) the properties of the output boundary state. Let us also stress that, while we rely on a specific representation of our quantum states in order to define the bulk/boundary map, our calculations and results could be reproduced in any other basis, with the disadvantage of a less transparent separation of bulk and boundary degrees of freedom. 

Next, we analyse the properties of the map so defined, and specifically we consider the conditions necessary for the map to be an isometry. We focus on a specific class of quantum states: the ones constructed out of individually weighted vertices \cite{Colafranceschi:2020ern}, whose wavefunctions are randomly distributed; the quantum properties of such states, including the isometry of the associated bulk-to-boundary map, are thus fully determined by the combinatorial structure of their entanglement graph, the dimension of the degrees of freedom attached to it and the probability distribution of the vertex wavefunctions. To study these class of states we exploit the random tensor networks techniques employed in \cite{Hayden:2016cfa}, adapting them to our context and generalizing them to account for the more general type of tensor networks our states correspond to.
	\noindent
	
	The first case corresponds to quantum states with fixed assignment of spin labels. This can be of two types, homogeneous (i.e. all spin labels are equal, the case considered in \cite{Hayden:2016cfa}) or inhomogeneous, with arbitrary assignment.  We find that, as a general condition, the map is isometric if and only if the reduced bulk state maximises the entropy. 
	Then we show that the bulk-to-boundary map of a homogeneous graph made of 4-valent vertices, each of them with at most one boundary link, cannot be isometric. 
Next, we show that the bulk-to-boundary map of a generic graph made of 4-valent vertices, each of them with at most one boundary link and spins pairwise equal, cannot be isometric. 
	Finally, for a generic graph made of 4-valent vertices, increasing the range of possible spins, i.e. the  \virg{inhomogeneity} of the assignment, increases the holographic character of the map.

\noindent The last case we consider corresponds to quantum states involving a superposition of spin labels for given graph. Here, the same general condition for isometry of the bulk-to-boundary map applies, and we write it explicitly in terms of the entanglement entropy of the reduced bulk state. In order to apply random tensor network techniques, and because we are anyway interested in quantum states with specific entanglement properties, we do not use the most general form of quantum gravity states as linear combinations of tensor networks but we work with states which are (generalised) tensor network themselves.

The structure of the paper is as follows:
in section \ref{EG}, we introduce the general quantum gravity framework we work in. In particular, we define the GFT field quanta with their spin network representation; we illustrate the gluing procedure for spin network vertices as a projection into maximally entangled states of open edges (leading to the possibility of understanding graph states as PEPS tensor networks); and we show in more detail the structure of graph states and the partition of their degrees of freedom into bulk and boundary ones.
In section \ref{map}, we explain how every graph state defines a map between bulk and boundary spaces; looking at the possible correlation between the degrees of freedom associated to such spaces (specifically, edge spins and intertwiners) we differentiate between the two cases corresponding to  \textit{fixed-spins assignment} and \textit{general sum over all possible spins}.
In section \ref{iso}, we study the isometry of bulk-to-boundary maps defined by \textit{fixed-spins} graph states. Similarly to \cite{Hayden:2016cfa}, we find that the maps are isometric if the corresponding reduced bulk states are maximally mixed. We check this condition for random graph states with uniform probability distribution for the vertex wavefunctions, by computing the average second order Rényi entropy. 

In section \ref{entropy}, we present some mathematical tools (generalization of that in \cite{Hayden:2016cfa}) that we use to investigate the properties of the bulk-to-boundary maps, in particular,  the computation of the second order Rényi entropy of a region of the bulk and/or of the boundary, for \textit{uniform probability distribution of the vertex wavefunctions}. We explain that the calculation of the average entropy is mapped to the partition functions of a classical Ising model defined on the graph itself.

In the last section, we discuss the more general case in which we superpose quantum states with different spin assignments, and the isometry condition in this case.

We conclude with a summary of the results and an outlook to future developments.

\section{Quantum gravity entanglement graphs}
\label{EG}

\subsection{GFT quanta: fundamental simplices dual to spin network vertices}
We consider quantum geometric GFT models \cite{Oriti:2011jm} in which field quanta are $(d-1)$-simplices, dual to $d$-valent vertices having edges decorated by $SU(2)$ variables, which are invariant under global $SU(2)$ action (gauge symmetry). Each edge is identified by a colour $i=1,...,d$ and denoted by $e^i$. The quantum-geometry state of a $d$-valent vertex is thus described by a function $f(g^1,...,g^d)$, where $g^i\in SU(2)$ is the group variable attached to the $i$-th edge $e^i$, which satisfies $f(g^1,...,g^d)= f(hg^1,...,hg^d)$ for every $h\in SU(2)$, i.e. the single-vertex Hilbert space is given by $\h_v=L^2(G^d/G)$. In the following, we use the vector notation to refer to the set of variables attached to the edges of a vertex, e.g. $\vec{g}\coloneqq g^1,...,g^d$.

By the Peter–Weyl theorem a function on ($d$ copies of) a compact group $G$ can be decomposed into irreducible representations of the latter. For $G=SU(2)$ this yields
\begin{equation}\begin{split}
&f(\vec{g})=\sum_{\vec{j}}\sum_{\vec{m}\vec{n}}  f^{\vec{j}}_{\vec{m}\vec{n}}\prod_i d_{j^i} D^{j^i}_{m^i n^i}(g^i),
\end{split}
\end{equation}
where $j^i\in \mathbb{N}/2$ are irreducible representations of $SU(2)$; the indices $m^i$ ($n^i$) label a basis in the vector space $V^{j^i}$ (its dual ${V^{j^i}}^*$) carrying the  representation $j^i$, which has dimension $d_{j}\coloneqq2j+1$; and $D^{j^i}_{m^i n^i}(g^i)$ is the matrix representing the group element $g^i$. Therefore, in this basis each edge $e^i$ is equipped with a representation spin $j^i$ and related magnetic numbers $m^i$ and $n^i$, which can be though of as associated to the edge endpoints (or, equivalently, to \virg{semi-edges}). After imposing the gauge invariance at each vertex, the vertex wavefunction takes the form 
\begin{equation}\begin{split}\label{stensor}
f(\vec{g})=\sum_{\vec{j}\vec{n}\iota} f^{\vec{j}}_{\vec{n}\iota} ~ \psi^{\vec{j}}_{\vec{n}\iota}(\vec{g}),
\end{split}
\end{equation}
where $\psi^{\vec{j}}_{\vec{n}\iota}(\vec{g})$ is the spin-network basis function:
\begin{equation}\label{spinn}
\psi^{\vec{j}}_{\vec{n}\iota}(\vec{g})\coloneqq\sum_{\vec{p}}C^{\vec{j} }_{\vec{p}\iota} \prod_i \sqrt{d_{j^i}} D^{j^i}_{p^i n^i}(g^i),
\end{equation}
with $C^{\vec{j}}_{\vec{p}\iota}\in \mathrm{Inv}_{SU(2)}\left[V^{j^1}\otimes ... \otimes V^{j^d}\right] \coloneqqrev \mathcal{I}^{\vec{j}}$ the intertwiner tensor. The Hilbert space of a vertex $v$ thus decomposes, via the Peter–Weyl theorem, as follows:
\begin{equation}\label{hv}
\mathcal{H}_v = \bigoplus_{\vec{j}_v}\left(\mathcal{I}^{\vec{j}_v}\otimes \bigotimes_{i=1}^dV^{j_v^i}\right)
\end{equation}
 We refer to the cited references on group field theory models and canonical loop quantum gravity for further details.

\subsection{Entanglement graphs and their degrees of freedom}

Two vertices are glued together by entangling their (semi-)edge degrees of freedom. In particular, given two vertices $v$ and $w$, connecting them along their $i$-th edges $e_v^i$ and $e_w^i$ corresponds to project their state into a maximally entangled state of $e_v^i$ and $e_w^i$ (for simplicity, we restrict the attention to a given edge spin $j$):
\begin{equation}\label{link}
\ket{e_{vw}^i}\coloneqq\frac{1}{\sqrt{d_{j}}}\sum_n  \ket{j n }\otimes \ket{jn} \in V^{j_v^i=j} \otimes V^{j_w^i=j}
\end{equation}
where $e_{vw}^i$ denotes the link connecting $v$ and $w$, carrying the edge colour $i$. From this perspective, GFT states associated to graphs can be seen as generalised tensor network structures: (symmetric) projected entangled pair states (PEPS) \cite{Orus:2013kga,Tagliacozzo_2011,Tagliacozzo_2014,Qi_2017}, with single-vertex tensors of the form $f^{\vec{j}}_{\vec{n}\iota}$ defined in \eq{stensor}. Since connectivity is determined by entanglement relations, we refer to such quantum gravity states also as \textit{entanglement graphs}.

The connectivity of a network/graph made of a set $V=\{1,...,N\}$ of vertices is encoded in its $N\times N$ adjacency matrix $A$~\cite{RevModPhys.42.271}, whose entries register the presence or absence of links between vertices: $A_{vw}=1$ if vertices $v$ and $w$ are connected by a link, and $A_{vw}=0$ otherwise. This encoding of network connectivity can be generalised to include the information of which (magnetic) indices of the vertex tensors are contracted: the components $A_{vw}$ are promoted to $d\times d$ matrices (where $d$ is the rank of tensors attached to the vertices), with the generic component $(A_{vw})_{ij}\coloneqq A_{(v-1)\cdot d+i,(w-1)\cdot d+j} $ equal to 1 if vertices $v$ and $w$ are connected along the $i$-th and $j$-th indices, respectively, and 0 otherwise. In the following, we assume that vertices can be connected only along edges of the same colour\footnote{In the tensorial group field theory formalism, the addition of coloring and the consequent restrictions on the allowed combinatorics of states and interactions is the key ingredient allowing control over their topology and, among many other results, the definition of large-N approximations~\cite{Gurau:2011xp,Bonzom:2011zz}.}. In terms of encoding of graphs into adjacency matrices, this implies $(A_{vw})_{ij}=0$ for all $i\neq j$. We denote by $L$ the set of internal links of the graph, i.e. $L = \{e_{vw}^{i} | A_{(v-1)\cdot d+i, (w-1)\cdot d+i}=1\}$, by $\partial \gamma$ the set of boundary edges, i.e. $\partial \gamma = \{e_v^i| A_{(v-1)\cdot d+i, (w-1)\cdot d+i}=0 ~ \forall ~ w \in V\}$, and by $E$ the set of all edges of the graph: $E=L \cup \partial \gamma$.

A generic state associated to a graph $\gamma$ of $N$ vertices can be constructed from a $N$-particle state $\ket{\varphi}\in \h_V\coloneqq \otimes_{v=1}^N \h_v$, by gluing vertices according to the combinatorial pattern of $\gamma$:
\begin{equation}\begin{split}
\ket{\varphi_\gamma}\coloneqq&\left(\bigotimes_{e^i_{vw}\in L} \bra{e_{vw}^i}\right)\ket{\varphi}\\=&\bigoplus_{\vec{j}_1...\vec{j}_N} \sum_{\vec{n}_1...\vec{n}_N}\sum_{\iota^1...\iota^N}\varphi^{\vec{j}_1...\vec{j}_N}_{\vec{n}_1...\vec{n}_N\iota_1...\iota_N}\prod_{e^i_{vw}\in L}\delta_{j_v^i,j_w^i}\delta_{n_v^i,n_w^i} \bigotimes_v \ket{\vec{j}_v \vec{n}_v \iota_v}
\end{split}
\end{equation}
 where $\ket{\vec{j}\vec{n}\iota}$ are spin network basis states with $\langle \vec{g}|\vec{j}\vec{n}\iota\rangle \coloneqq \psi^{\vec{j}}_{\vec{n}\iota}(\vec{g})$ defined in \eq{spinn}; the Kronecker deltas impose maximal entanglement between semi-edge degrees of freedom, as a result of the projection of the $N$-particle state $\ket{\varphi}$ into the link states defined in \eq{link}. The resulting graph wavefunction $\varphi_\gamma$ thus takes the form
\begin{equation}\label{lwf}
{\varphi_\gamma}^{\{j_{e\in E}\}}_{\{n_{e\in \partial \gamma}\}\{\iota_{v\in V}\}}=\varphi^{\vec{j}_1...\vec{j}_N}_{\vec{n}_1...\vec{n}_N\iota_1...\iota_N}\prod_{e^i_{vw}\in L}\delta_{j_v^i,j_w^i}\delta_{n_v^i,n_w^i} 
\end{equation}
with $j_e$ ($n_e$) spin (magnetic number) associated to edge $e$. States with an entanglement pattern $\gamma$ live in a subspace of $\h_V\coloneqq \otimes_{v=1}^N \h_v$ given by
\begin{equation}\label{h}
\mathcal{H}_\gamma=\bigoplus_{J} \left(\bigotimes_v\mathcal{I}^{\vec{j}_v}\otimes \bigotimes_{e\in \partial \gamma}V^{j_e}\right)
\end{equation} 
 where $J$ is a set of spins associated to the graph edges: $J=\{j_e| e \in E\}$.

Note that, in general, the $N$-particle wavefunction $\varphi$ does not factorize over single-vertex states (though it can be expanded in the basis of open spin network vertices). In the special case in which $\varphi$ features such a factorization, its spin network expansion takes the following form: \begin{equation}\label{fact}
	\varphi^{\vec{j}_1...\vec{j}_N}_{\vec{n}_1...\vec{n}_N\iota_1...\iota_N}=\prod_{v=1}^{N}(f_v)_{\vec{n}_v\iota_v}^{\vec{j}_v}
	\end{equation}
where $(f_v)_{\vec{n}_v\iota_v}^{\vec{j}_v}$ is the wavefunction associated to vertex $v$. In our analysis we will focus on graph states constructed out of many-body wavefunctions having this particular form; however, for the present discussion we do not assume any specific form of $\varphi$, which is therefore completely generic. 

Let us now focus on the degrees of freedom of a graph state $\varphi_\gamma$; they are the following:
\begin{itemize}
\item[a.] spins $j_v^i$ and magnetic numbers $n_v^i$ associated to the boundary edges $e_v^i \in \partial \gamma$;
\item[b.] spins $j_{vw}^i$ associated to the internal links  $e_{vw}^i \in L $;
\item[c.]  intertwiner quantum numbers $\iota_v$ associated to the vertices $v\in V$, collectively indicated as $\dot{\gamma}$.
\end{itemize}
The set $a$ corresponds to the \textit{boundary degrees of freedom}, while the sets $b$ and $c$ identify the \textit{bulk degrees of freedom}; in particular, the set $b$ contains information on the combinatorial structure of the bulk and the dimension of the internal links, while $c$ can be interpreted as a set of \virg{internal} degrees of freedom anchored to the vertices. From the simplicial-geometry perspective, in fact, the intertwiner labels determine the volume of the simplices dual to the graph vertices, while the spin labels carry information about areas of surfaces, dual to the graph edges, which can be in the bulk or in the boundary. Since $c$ is not independent from $a$ and $b$, the graph Hilbert space does not factorize into bulk and boundary Hilbert spaces. However, as can be seen from \eq{h} such a factorization takes place in every fixed-spins subspace. In the next section, we show that fixed-spins graph states naturally define maps between their bulk and boundary degrees of freedom, and that such a feature extends to completely generic (i.e. involving spin-superposition) graph states, upon embedding the graph Hilbert space $\h_\gamma$ into the tensor product of generalized bulk/boundary Hilbert spaces.
 \section{Entanglement graphs as bulk-to-boundary maps}\label{map}
In this section we show how one can use the quantum gravity states to define maps between their bulk and boundary degrees of freedom, as defined above; for closely related recent work, see \cite{Chen:2021vrc}. Although this feature holds for any graph state, we specialize to quantum states with an almost factorized form of the wavefunction, i.e. with individual vertex contributions subject only to the entanglement contractions corresponding to the combinatorial pattern of the graph to which it is associated. Therefore, our quantum states are themselves tensor networks, rather than generic linear combinations of them (which is true for any quantum state in the GFT formalism). This will make more explicit the possibility to regard entanglement graphs as maps on a bipartition of their degrees of freedom, similarly to what happens for tensor network states/codes \cite{Hayden:2016cfa}. 

\subsection{Bulk and boundary subspaces}
To better clarify the splitting of an entanglement-graph degrees of freedom into bulk and boundary ones, we first illustrate it for the basic graph structure: a single spin-network vertex, whose Hilbert space can be written as
\begin{equation}
\mathcal{H}_v = \bigoplus_{\vec{j}_v} \mathcal{H}_v(\vec{j}_v), \qquad \mathcal{H}_v(\vec{j}_v)=\mathcal{I}^{\vec{j}_v}\otimes \bigotimes_{i=1}^dV^{j_v^i}
\end{equation}
Every fixed-spins subspace  $\mathcal{H}_v(\vec{j}_v)$ is given by the tensor product of the intertwiner space $\mathcal{I}^{\vec{j}_v}$, describing the \textit{bulk} degree of freedom, and the space associated to the open lines of the vertex, namely its \textit{boundary}: $\bigotimes_{i=1}^dV^{j_v^i}$. In every $\mathcal{H}_v(\vec{j}_v)$ we can thus consider a factorization of the spin-network basis $\ket{\vec{j}\vec{n}\iota}$ into the bulk component $\ket{\vec{j}\iota}$ and the boundary one $\ket{j^1 n^1}\otimes ... \otimes \ket{j^d n^d}$.Table~\ref{tab:ver} summarises Hilbert space and corresponding basis of bulk and boundary degrees of freedom of a spin-network vertex with fixed spins.
\begin{table}[ht]
	\centering
	
	\renewcommand{\arraystretch}{3}
	\begin{tabular}{>{\centering}m{0.9in}|>{\centering}m{1.5in}| >{\centering\arraybackslash}m{1.2in}}
		\toprule
		Object& Hilbert space& Basis state \\
		\midrule
		boundary&	$V^{j^1}\otimes ... \otimes V^{j^d}$ &$\ket{j^1n^1}\otimes...\otimes\ket{j^dn^d}$ \\ bulk (intertwiner) & $\mathcal{I}^{\vec{j}}= \text{Inv}_{SU(2)}\left[\bigotimes_i V^{j^i}\right]$&
		$\ket{\vec{j}~\iota}$ \\
		\bottomrule
	\end{tabular}
	\caption{Fixed-spins subspace and corresponding basis of the vertex substructures.}
	\label{tab:ver}
\end{table}

Consider now a set of $N$ vertices described by wavefunctions $\{f_v\}_{v=1}^N$ picked on edge spins $\{\vec{j}_v\}_{v=1}^N$, and a graph $\gamma$ (i.e. a combinatorial pattern $A$) compatible with them (specifically, with their edge spins, which must coincide for edges to be glued). A state associated to $\gamma$ can be obtained by contracting $\bigotimes_v \ket{f_v}$ with link states, specifically a state $\ket{e_{vw}^i}$ for each adjacency-matrix element $(A_{vw})_{ij}=1$, where $\ket{e_{vw}^i}$ is a maximally entangled state of $e_v^i$ and $e^{w}_i$ (see \eq{link}):
\begin{equation}\begin{split}
\label{phigamma}
\ket{\phi_\gamma} \coloneqq &
\left(\bigotimes_{e^i_{vw}\in L} \bra{e_{vw}^i}\right)\bigotimes_v \ket{f_v}\\=&\sum_{\{n_{e \in \partial\gamma}\}}\sum_{\iota_1,...,\iota_N} \left(\phi_\gamma\right)_{\{n_{e \in \partial\gamma}\}\iota_1,...,\iota_N} \bigotimes_{e\in \partial \gamma}\ket{j_e n_e}\otimes \bigotimes_v \ket{\vec{j}_v~\iota_v}
\end{split}
\end{equation}
with
\begin{equation}\label{phiJ}
\left(\phi_\gamma\right)_{\{n_{e \in \partial\gamma}\}\iota_1,...,\iota_N}= \sum_{\{n_{e \in L}\}} \sum_{\{p\}}  \prod_v(f_v)_{\vec{n}_v\iota_v}^{\vec{j}_v} \prod_{e_{vw}^i\in L} \delta_{n_v^i p^{vw}_i}\delta_{n_w^i p^{vw}_i}
\end{equation}
where $\{n_{e \in L}\}$ is the set of magnetic indices associated to the internal semi-links. Note that $ \bigotimes_v \ket{\vec{j}_v~\iota_v}$ is the basis element of the Hilbert space describing the \textit{bulk} of the entanglement graph, namely the set of intertwiners attached to the vertices:
\begin{equation}\label{bulk}
\mathcal{H}_{\dot{\gamma}}(J) \coloneqq \bigotimes_v\mathcal{I}^{\vec{j}_v},
\end{equation}
where $J$ is the set of spins attached to the edges of $\gamma$, while $\bigotimes_{e\in \partial \gamma}\ket{j_e n_e}$ is the basis element of the Hilbert space associated to the \textit{boundary}, i.e.  
\begin{equation}\label{boundary}
\h_{\partial\gamma}(J_\partial)\coloneqq \bigotimes_{e\in \partial \gamma}V^{j_e},
\end{equation}
where $J_\partial=\{j_e| e \in \partial \gamma\}\subset J$ is the set of spins attached to the boundary edges.

The above example shows that the Hibert space $\mathcal{H}_\gamma(J)$ associated to a fixed-spins entanglement graph (of which \eq{phigamma} is a particular case) factorizes into bulk and boundary spaces: $\mathcal{H}_\gamma(J)=\h_{\partial\gamma}(J_\partial)\otimes \mathcal{H}_{\dot{\gamma}}(J)$. Note that $\mathcal{H}_\gamma(J)$ is just the $J$-subspace of the Hilbert space $\h_\gamma$ describing a generic (i.e. involving superposition of spins) entanglement graph, defined in \eq{h}; in fact, $\h_\gamma=\bigoplus_{J} \mathcal{H}_\gamma(J)$.

\subsection{Fixed-spins entanglement graphs as bulk-to-boundary maps}
 Consider the PEPS constructed from vertex states $\{f_v\}_{v=1}^N$ picked on edge spins $\{\vec{j}_v\}_{v=1}^N$,  according to the combinatorial pattern $\gamma$, i.e. the state $\ket{\phi_\gamma}$ of \eq{phigamma}. Given a state in the bulk Hilbert space of \eq{bulk},  
\begin{equation}\label{input}
\ket{\zeta}=\sum_{\iota_1,...,\iota_N} \zeta_{\iota_1,...,\iota_N}\bigotimes_v \ket{\vec{j}_v~\iota_v},
\end{equation}
we can construct a corresponding boundary state as follows:
\begin{equation}\begin{split}\label{output}
\ket{\phi_{\partial \gamma}(\zeta)} &\coloneqq  \bra{\zeta}\phi_{\gamma}\rangle\\&=\bra{\zeta} \left(\bigotimes_{e^i_{vw}\in L} \bra{e_{vw}^i}\right)\bigotimes_v \ket{f_v}\\&=\sum_{\{n_{e\in\partial\gamma}\}}  \left(\phi_{\partial \gamma}(\zeta)\right)_{\{n_{e\in\partial\gamma}\}} \bigotimes_{e\in \partial \gamma}\ket{j_e n_e}
\end{split}
\end{equation}
with 
\begin{equation}\small
\left(\phi_{\partial \gamma}(\zeta)\right)_{\{n_{e\in\partial\gamma}\}} =\sum_{\{\iota\}}\sum_{\{n_{e\in L}\}} \sum_{\{p\}} \zeta^*_{\iota_1,...,\iota_N} \prod_v(f_v)_{\vec{n}_v\iota_v}^{\vec{j}_v} \prod_{e_{vw}^i\in L} \delta_{n_v^i p^{vw}_i}\delta_{n_w^i p^{vw}_i}.
\end{equation}
Now, \eq{input} and \eq{output} can be considered, respectively, \emph{input} and \emph{output} of a map defined between the bulk and the boundary Hilbert spaces, i.e.
\begin{equation}\begin{split}
M[\phi_\gamma]~:&~ \mathcal{H}_{\dot{\gamma}}(J) \coloneqq \bigotimes_v\mathcal{I}^{\vec{j}_v}~\rightarrow~\h_{\partial\gamma}(J_\partial)\coloneqq \bigotimes_{e\in \partial \gamma}V^{j_e},
\end{split} 
\end{equation}
which acts as follows:
\begin{equation}\begin{split}
M[\phi_\gamma]\ket{\zeta}=\langle \zeta |\phi_\gamma\rangle=\ket{\phi_{\partial \gamma}(\zeta)}.
\end{split}
\end{equation}
Therefore, the entanglement graph $\ket{\phi_\gamma}$ naturally defines a map $M[\phi_\gamma]$ from the bulk to the boundary degrees of freedom, whose action amounts to feeding the bulk with an input state that establishes correlations among the intertwiners, and returning the corresponding boundary state. 

Note that, between the same bulk and boundary Hilbert spaces, we have a family of similar maps $M[\varphi_{\gamma'}]$ defined by all possible states $\ket{\varphi_{\gamma'}}$ associated to all possible graphs $\gamma'$ having the same bulk and boundary of $\gamma$ (i.e. same number of vertices and same boundary edges), with action $M[\varphi_{\gamma'}]\ket{\zeta}=\langle \zeta |\varphi_{\gamma'}\rangle$.

\subsection{Generic entanglement graphs as bulk-to-boundary maps}\label{genericmap}
In the more general case of vertex wavefunctions $\{f_v\}_{v=1}^N$ that spread over all possible values of the edge spins, a (generalised) PEPS with combinatorial pattern $\gamma$ takes the form
\begin{equation}\begin{split}\label{phi}\small
\ket{\phi_\gamma}&= \left(\bigotimes_{e^i_{vw}\in L} \bra{e_{vw}^i}\right)\bigotimes_v \ket{f_v}\\&=\bigoplus_{J}\sum_{\{\iota\}} \sum_{\{n_{e \in \partial\gamma}\}} \left(\phi_\gamma^J\right)_{\{n_{e \in \partial\gamma}\}\iota_1...\iota_N} \bigotimes_{e\in \partial \gamma}\ket{j_e n_e}\otimes \bigotimes_{v }\ket{\vec{j}_v\iota_v}
\end{split}
\end{equation}
with 
\begin{equation}\label{edge}
\ket{e_{vw}^i}=\bigoplus_j\frac{1}{\sqrt{d_{j}}}\sum_n  \ket{jn}\otimes \ket{jn}.
\end{equation}
and $\left(\phi^J_\gamma\right)_{\{n_{e \in \partial\gamma}\}\iota_1...\iota_N}$ having the form of the fixed-spins graph-wavefunction of \eq{phiJ}.
Given a state $\ket{\zeta}$ for a set of $N$ intertwiners recoupling arbitrary spins, 
\begin{equation}\label{z}
\ket{\zeta}=\bigoplus_{J}\sum_{\iota_1,...,\iota_N} \left(\zeta^J\right)_{\iota_1,...,\iota_N}\bigotimes_v\ket{\vec{j}_v~\iota_v},
\end{equation}
the corresponding boundary state is
\begin{equation}\begin{split}\label{out}
\ket{\phi_{\partial \gamma}(\zeta)} &=  \bra{\zeta}\phi_\gamma \rangle \\&=  \bra{\zeta}\left(\bigotimes_{e^i_{vw}\in L} \bra{e_{vw}^i}\right)\bigotimes_v \ket{f_v}\\&=\bigoplus_{J_{\partial }}\sum_{\{n_{e\in \partial\gamma}\}} \left(\phi_{\partial \gamma}^{J_{\partial }}(\zeta)\right)_{\{n_{e \in \partial\gamma}\}}  \bigotimes_{e\in \partial \gamma}\ket{j_e n_e}
\end{split}
\end{equation}
with 
\begin{equation}\begin{split}
\left(\phi_{\partial \gamma}^{J_{\partial }}(\zeta)\right)_{\{n_{e \in \partial\gamma}\}}  =&\sum_{J_L}\sum_{\{\iota\}}\sum_{\{n_{e \in L}\}} \sum_{\{p\}} \left(\zeta^J\right)^*_{\iota_1...\iota_N} \prod_v(f_v)_{\vec{n}_v\iota_v}^{\vec{j}_v} \prod_{e_{vw}^i\in L} \delta_{n_v^i p^{vw}_i}\delta_{n_w^i p^{vw}_i}\delta_{j_v^i j_{vw}^i}\delta_{j_w^i j_{vw}^i}
\end{split}
\end{equation}
where $J_L=\{j_{vw}^i|e_{vw}^i \in L\}$ is the set of spins attached to internal links.  Note that, despite the bulk state being associated to a set of intertwiners among arbitrary spins, the contraction with $\ket{\phi_\gamma}$ returns only those configurations which are compatible with the combinatorial pattern $\gamma$.

We can regard \eq{z} and \eq{out} as, respectively, \textit{input} and \textit{output} of a map $M[\phi_\gamma]$ acting between the following (generalised) bulk and boundary spaces (associated to  entanglement graphs having combinatorial pattern $\gamma$):
\begin{equation}\begin{split}\label{spaces}
\h_{\dot{\gamma}} = \bigoplus_{J}\h_{\dot{\gamma}}(J), \qquad \h_{\partial\gamma}=\bigoplus_{J_{\partial}}\h_{\partial\gamma}(J_\partial)
\end{split}
\end{equation}
Note that the graph states with combinatorial pattern $\gamma$ live in a particular subspace of $\h_{\dot{\gamma}}  \otimes \h_{\partial\gamma}$, the one which contains the connectivity-induced correlation between spins of the boundary and intertwiners of the bulk. Note also that the output boundary states belong to the subspace of $\h_{\partial\gamma}$ defined by the gauge symmetry on set of links pertaining to the same vertex.

By considering the most general bulk and boundary Hilbert spaces, we are thus able to regard \textit{every} graph state $\varphi_{\gamma}$ (involving an arbitrary spin superposition) as a map between them. The correlation between bulk and boundary degrees of freedom in $\varphi_{\gamma}$ translates into a correlation between the input and output subspaces connected by the map; in particular, the only non-vanishing components of the map are the following:
\begin{equation}\label{comp}
\left(\bigotimes_{e\in \partial \gamma}\bra{j_e n_e}\right)M[\varphi_{\gamma}]\left(\bigotimes_v\ket{\vec{j}_v~\iota_v}\right)
\end{equation}
where $J_{\partial \gamma}\subset J$.

Let us remark that, despite the possibility of reading a generic graph state as a bulk-to-boundary map in the sense specified above, the graph Hilbert space itself cannot be factorized into boundary and bulk spaces due to the sharing of degrees of freedom between these two structures (the intertwiner degree of freedom depends on the incident spins). Note however that, in order to make the bulk degrees of freedom independent from the boundary ones, it is sufficient to fix the spins $J_{\partial}$ of the boundary. The full bulk Hilbert space $\h_{\dot{\gamma}}$ is then redundant, as it actually reduces to a direct sum of $J$-subspaces such that the boundary portion of $J$ coincides with $J_\partial$.

\section{Isometry condition for bulk-to-boundary maps (fixed-spins case)}\label{iso}

We proceed to determine the conditions under which the bulk-to-boundary map associated to a graph state is isometric. We restrict the attention to states of the form $\ket{\phi_\gamma}=\bigotimes_{e\in\gamma} \bra{e}\bigotimes_v \ket{f_v}$, although the analysis can be easily extended to arbitrary states.Here we focus on the fixed-spins case, where all vertex wavefunctions $\{f_v\}_v$ are picked on specific values $\{\vec{j}_v\}_{v}$ of the edge spins, and address the spin superposition case in Section \ref{isosum}. 
	
As explained in Section \ref{map}, the state $\ket{\phi_\gamma}$ can be thought of as a bulk-to-boundary map $M[\phi_\gamma]$
acting on an a bulk basis state $\bigotimes_v\ket{\vec{j}_v~\iota_v}$ as follows:
\begin{equation}\begin{split}\label{M}
M[\phi_\gamma]\bigotimes_v\ket{\vec{j}_v~\iota_v}=\left(\bigotimes_v\bra{\vec{j}_v~\iota_v}\right) \ket{\phi_\gamma}
\end{split}
\end{equation} 
To simplify the map notation, in the following we omit for $M[\phi_\gamma]$ the explicit reference to the state from which it is defined. The map $M$ is isometric if and only if it satisfies\begin{equation}
	M^\dagger M = \mathbb{I},
\end{equation}
where $\mathbb{I}$ is the identity operator. We are going to show that this condition is equivalent to the bulk reduction of the graph state being completely mixed.

\begin{figure}[t]
	\centering
	\includegraphics[width=0.45\linewidth]{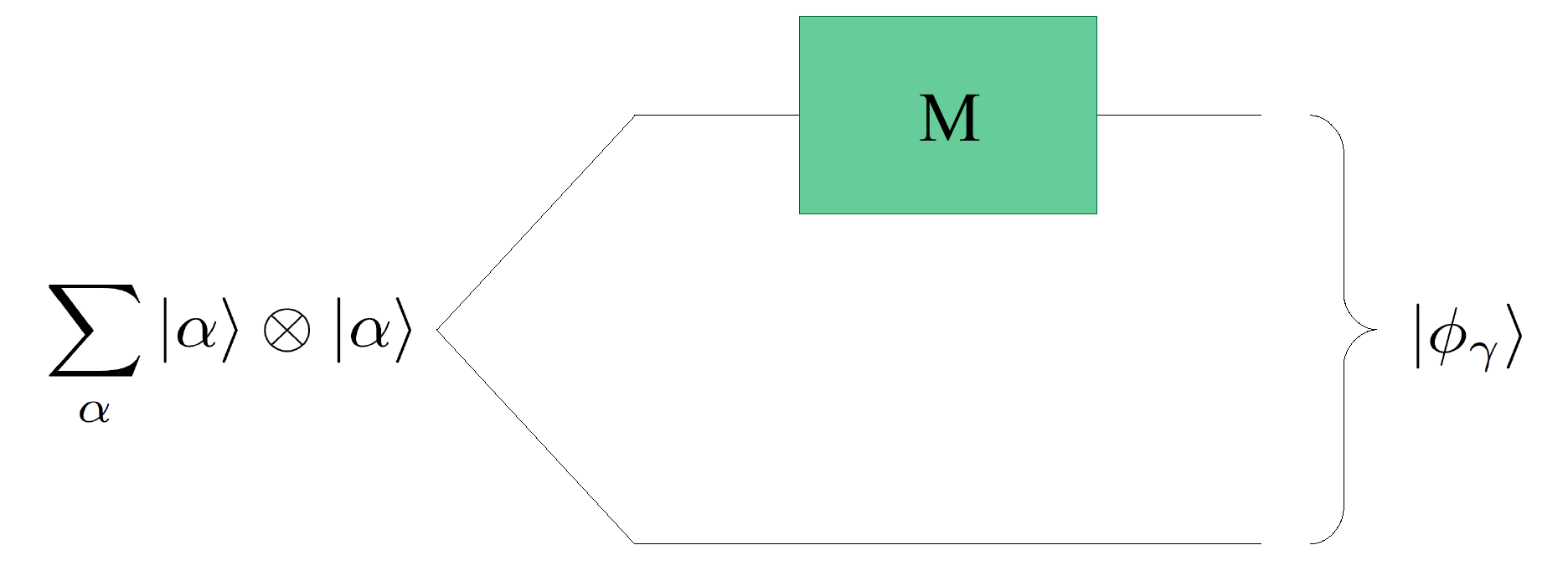}\caption{The graph state $\ket{\phi_\gamma}$ can be obtained starting from two copies of the bulk in a maximally entangled state, and applying to one of them the bulk-to-boundary map $M$ defined  in \eq{M}. When the reduced bulk state is maximally mixed, this qualifies as Choi-Jamio\l kowski isomorphisms between the state  $J(M)\coloneqq\frac{1}{D_{\dot{\gamma}}} |\phi_\gamma\rangle \langle \phi_\gamma|$ and the quantum channel $M$.}
	\label{channel}
\end{figure}

Let $\ket{\alpha}$ be a short notation for the basis element of the bulk Hilbert space; in terms of the map $M$, the graph state $\ket{\phi_\gamma}$ can be written as follows:
\begin{equation}\label{choi}
\ket{\phi_\gamma}=\sum_{\alpha}\left(M \otimes \mathbb{I}\right) \ket{\alpha}\otimes \ket{\alpha}
\end{equation}
That is, $\ket{\phi_\gamma}$ is obtained by applying the map $M$ to a branch of two maximally entangled copies of the bulk, as depicted in figure \ref{channel}.
The normalised reduced bulk-state then takes the form
\begin{equation}\begin{split}
  \rho_{\dot{\gamma}}=\text{Tr}_{\partial \gamma} \left(\frac{|\phi_\gamma\rangle \langle \phi_\gamma|}{D_{\dot{\gamma}}}\right)=\frac{1}{D_{\dot{\gamma}}} \sum_{\alpha\alpha'} \left(M^\dagger  M \right)_{\alpha\alpha'}^*|\alpha\rangle\langle\alpha'|
\end{split}
\end{equation}
where $D_{\dot{\gamma}}=\prod_v D_{\vec{j}_v}$ is the bulk-space dimension. From this expression, it is immediate to realize that the isometry condition $M^\dagger M=\mathbb{I}$ translates into the requirement that $\rho_{\dot{\gamma}}$ is maximally mixed, i.e.
\begin{equation}\begin{split}\label{J}
\rho_{\dot{\gamma}}=\text{Tr}_{\partial \gamma} \left(\frac{|\phi_\gamma\rangle \langle \phi_\gamma|}{D_{\dot{\gamma}}}\right)=\frac{\mathbb{I}}{  D_{\dot{\gamma}} }
\end{split}
\end{equation}
This can be checked by verifying that $\rho_{\dot{\gamma}}$ has maximum entropy. 

Interestingly, the relation between $M$ and $\ket{\phi_\gamma}$ expressed in \eq{choi}, with $\ket{\phi_\gamma}$ satisfying \eq{J}, is known in quantum information theory as Choi-Jamio\l kowski isomorphisms \cite{WatrousNotes2011}, a dualism between quantum channels (i.e. completely positive trace-preserving maps) and quantum states. In particular, the (bulk-normalised) graph state $J(M)\coloneqq \frac{1}{D_{\dot{\gamma}}} |\phi_\gamma\rangle \langle \phi_\gamma|$, when satisfying \eq{J}, turns out to be the Choi-Jamio\l kowski state associated to the quantum channel $M$ (by reversing the point of view, the bulk-to-boundary map $M$ turns out to be a quantum channel when $\ket{\phi_\gamma}$ satisfies \eq{J}).

\subsection{Average bulk entropy for random entanglement graphs}\label{average}
 To determine the entropy content of our bulk state, we focus on the second order Rényi entropy. For a portion $P$ of a quantum system described by the density matrix $\rho$, this is defined as
\begin{equation}
S_2(\rho_P)=-\log \Tr\left(\rho_P^2 \right)
\end{equation}
where $\rho_P=\Tr_{\overline{P}}(\rho)$, with $\overline{P}$ the subsystem complementary to $P$. The computation of $S_2$ can be more easily performed by applying the replica trick:
\begin{equation}\label{swap}
S_2(\rho_{P})=-\log \left(\frac{\Tr \left[\left(\rho \otimes \rho\right)\mathcal{S}_P\right]}{\Tr\left[\rho \otimes \rho\right]}\right),
\end{equation}
where $\mathcal{S}_P$ is the swap operator acting on the two copies of subsystem $P$. In a more compact form, we have
\begin{equation}
S_2(\rho_{P})=-\log \left(\frac{Z_1}{Z_0}\right),
\end{equation}
where 
\begin{equation}\begin{split}
\label{z10}
&Z_1 \coloneqq \Tr \left[\left(\rho \otimes \rho\right)\mathcal{S}_P\right]\\&Z_0 \coloneqq \Tr \left[\rho \otimes \rho\right].
\end{split}
\end{equation}
We are interested in $S_2(\rho_{\dot{\gamma}})$, which according to \eq{swap} can be written as
\begin{equation}
S_2(\rho_{\dot{\gamma}})=-\log \left(\frac{\Tr \left[\left(\rho \otimes \rho\right)\mathcal{S}_{\dot{\gamma}}\right]}{\Tr\left[\rho \otimes \rho\right]}\right),
\end{equation}
where $\rho=\ket{\phi_\gamma}\bra{\phi_\gamma}$ and $\mathcal{S}_{\dot{\gamma}}$ is the swap operator for the whole bulk (for each vertex of the graph, it swaps the two copies of the intertwiner Hilbert space). 
The quantities $Z_1$ and $Z_0$ can then be expressed as follows:
\begin{equation}\begin{split}\label{z1z0}
&Z_{1/0}=  \Tr \left[\rho_L^{\otimes 2}\bigotimes_v \left(\ket{f_v}\bra{f_v}\right)^{\otimes 2}\mathcal{S}_{\dot{\gamma}}/\mathbb{I}\right],
\end{split}
\end{equation}
where $\rho_L \coloneqq \bigotimes_{e \in L} \ket{e}\bra{e}$, with $\ket{e}$ short notation for the maximally entangled state of the internal link $e$.

We consider the case in which every vertex wavefunction $f_v$ is chosen  independently at random from its Hilbert space, according to the uniform probability measure. Similarly to \cite{Hayden:2016cfa}, the tensor $(f_v)_{\vec{n} \iota}^{\vec{j}}$ associated to the vertex $v$ is regarded as a vector of dimension 
\begin{equation}\label{dim}
D_v(\vec{j})\coloneqq d_{j^1}\cdot ... \cdot d_{j^d} \cdot  D_{\vec{j}},
\end{equation}
and the computation of typical values of functions of $f_v$ is performed by setting  $\ket{f_v}=U \ket{f^0_v}$, where $\ket{f^0_v}$ is a reference state and $U\in SU(D_v)$ (when working in the fixed-spins setting, we omit from the vertex dimension of \eq{dim} the explicit reference to $\vec{j}$), and integrating over $U$ with the Haar measure. Regarding the combinatorial structure of the graph, we assume that every vertex has at most one boundary edge, and that vertices cannot share more than one link. 

\begin{figure}[t]
	\begin{minipage}[t]{0.46\linewidth}
		\centering
		\includegraphics[width=0.9\linewidth]{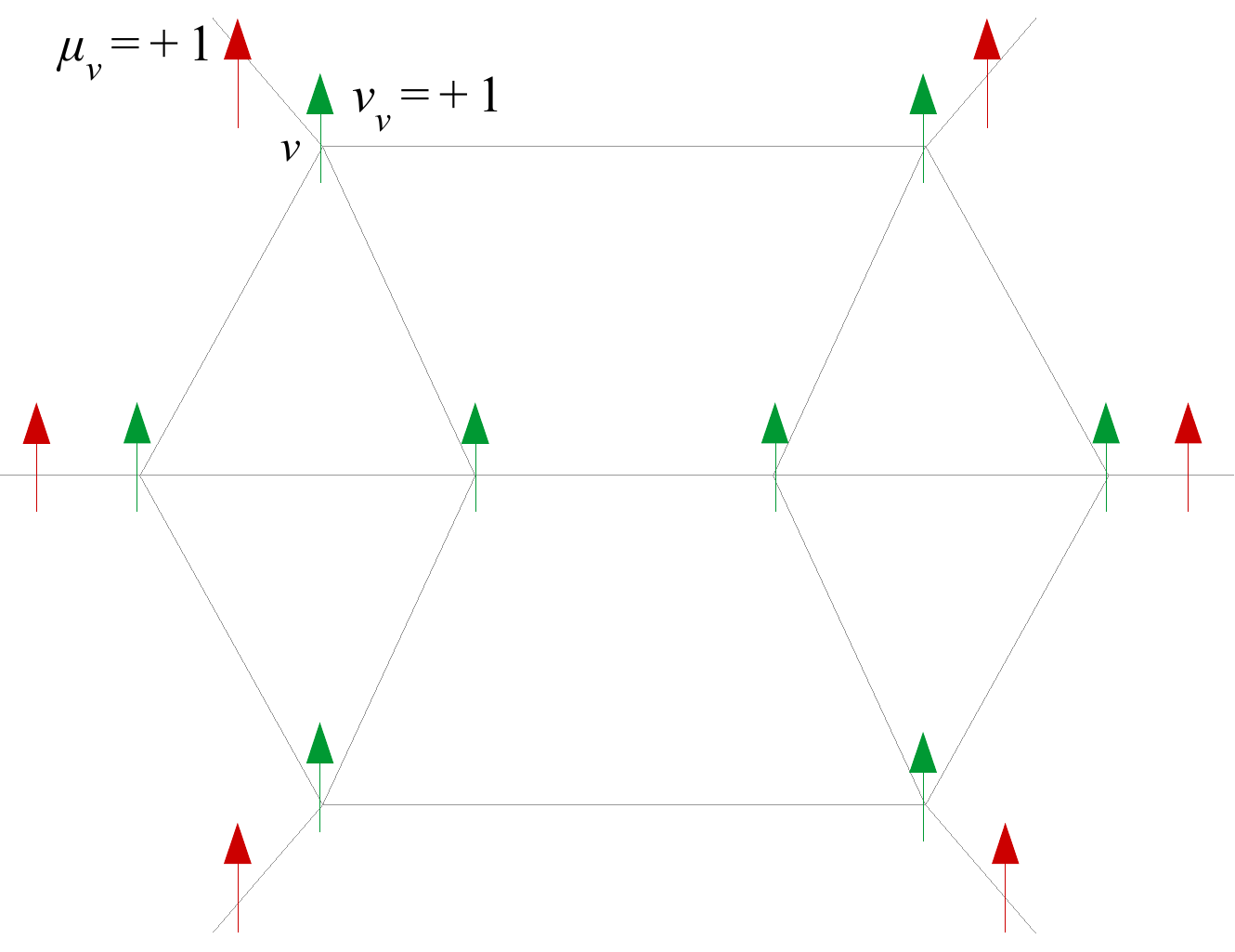}\caption{In the Ising action $\mathcal{A}_0$, both the boundary spins $\vec{\mu}$ (shown in red) and the bulk ones $\vec{\nu}$ (shown in green) point up: $\mu_v=+1$, $\nu_v=+1$ $\forall v$.}
		\label{F0}
	\end{minipage}
\hspace{0.5cm}
	\begin{minipage}[t]{0.46\linewidth}
		\centering
		\includegraphics[width=0.9\linewidth]{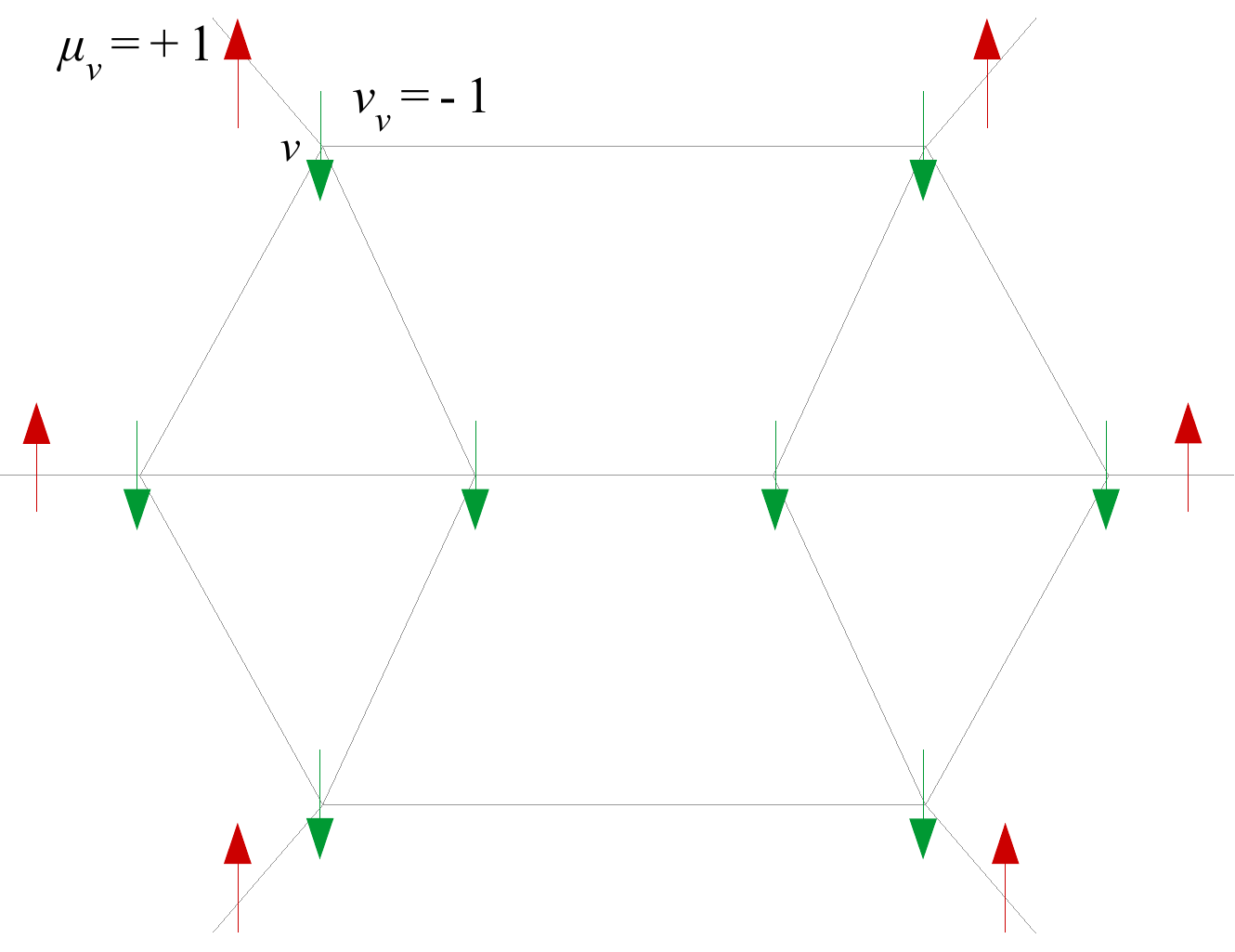}\caption{In the Ising action $\mathcal{A}_1$, the boundary spins (in red) point up, but the bulk spins (in green) are flipped down: $\mu_v=-\nu_v=+1$ $\forall v$.}
		\label{F1}
	\end{minipage}
\end{figure}

 Given the random character of the vertex wavefunctions, we focus on the case of large spins, in which deviations from typical values of the quantities we are interested in are significantly suppressed.  In this case, it holds that \cite{Hayden:2016cfa}
\begin{equation}\label{av}
\overline{S_2(\rho_{\dot{\gamma}})}\approx-\log \frac{\overline{Z_1}}{\overline{Z_0}},
\end{equation}
and the calculation of $\overline{S_2(\rho_{\dot{\gamma}})}$ then reduces to the evaluation of $\overline{Z_{1}}$ and $\overline{Z_{0}}$.
 As we explain in detail in Section \ref{entropy}, the quantities $\overline{Z_{1}}$ and $\overline{Z_{0}}$ are equivalent to partition functions of a classical Ising model defined on the graph $\gamma$. In particular, each vertex $v$ carries a Ising spin $\sigma_v$ arising from the randomization over its wavefunction. In addition to that, boundary edges carry spins $\mu_v$ (where $v$ refers to the source vertex), and vertices carry spins $\nu_v$ (also called \virg{bulk spins}, as they are related to the intertwiner degrees of freedom), whose value depends on whether ($-1$) or not ($+1$) they are part of the region respect to which the entropy is computed; in fact, these additional spins arise from the swap operator entering \eq{swap}. Specifically, $\overline{Z_{1}}$ and $\overline{Z_{0}}$ are particular cases of 
 \begin{equation}\label{Z}
     \overline{Z}(\vec{\mu}, \vec{\nu})=\sum_{\vec{\sigma}}e^{-\mathcal{A}[\vec{\mu}, \vec{\nu}](\vec{\sigma})}
 \end{equation}
where $\vec{\sigma}$ refers to the set of Ising spins, $\vec{\mu}$ ($\vec{\nu}$) to that of boundary (bulk) pinning fields, and
\begin{equation}\begin{split}\label{A}
\mathcal{A}[\vec{\mu}, \vec{\nu}](\vec{\sigma})\coloneqq& -\frac{1}{2}\left[\sum_{e_{vw}^i\in L}(\sigma_v\sigma_w -1)\log d_{j_{vw}^i}
+\sum_{e_v^i\in \partial \gamma}(\sigma_v\mu_v-1)\log d_{j_v^i} +\sum_{v}(\sigma_v\nu_v-1)\log D_{\vec{j}_v}\right] + k
\end{split}
\end{equation}
is an Ising-like action ($k$ is a constant term); the quantity $\overline{Z_{0}}$ ($\mathcal{A}_0$) is given by \eq{Z} (\eq{A}) with $\mu_v=\nu_v=+1$ for all $v$, while the quantity $\overline{Z_{1}}$ ($\mathcal{A}_1$) is given by \eq{Z} (\eq{A}) with $\mu_v= -\nu_v=+1$ for all $v$ (see Figure \ref{F0} and Figure \ref{F1}, respectively). In fact, as we are looking to the reduced bulk state, the swap operator in \eq{swap} acts only on the intertwiner Hilbert spaces, thus only the bulk spins $\vec{\nu}$ change sign between $\overline{Z_{0}}$ and $\overline{Z_{1}}$. Defining the free energies $F_{1}\coloneqq-\log \overline{Z_{1}}$ and $F_{0}\coloneqq-\log \overline{Z_{0}}$, the average entropy given by \eq{av} can be computed as
\begin{equation}\begin{split}\label{FF}
\overline{S_2(\rho_{\dot{\gamma}})}&\approx F_1 -F_0,
\end{split}
\end{equation}
namely as the energy cost of flipping down the bulk spins $\{\nu_v\}$. From \eq{Z} and \eq{A} it follows that in the case of large spins the partition function $\overline{Z}(\vec{\mu}, \vec{\nu})$ is dominated by the lowest energy configuration, so that 
\begin{equation}\label{approx}
F(\vec{\mu}, \vec{\nu}) = -\log \overline{Z}(\vec{\mu}, \vec{\nu}) \approx  \min_{\vec{\sigma}}\mathcal{A}[\vec{\mu}, \vec{\nu}](\vec{\sigma}) 
\end{equation} 
Note that, as far as the thermodynamic picture of the Ising model is concerned, the large spins scenario corresponds to the Ising temperature going to zero, and the system dropping to the lowest energy configuration.

In the following we consider some special cases of quantum states, more precisely of spin assignments, in order to gain insights on the behaviour of the entropy.

\subsubsection{Homogeneous case}
In the homogeneous case, i.e. with all edge spins equal to the same value $j$, the Ising action takes the form

\begin{figure}[t]
	\begin{minipage}[t]{0.46\linewidth}
		\centering
		\includegraphics[width=0.9\linewidth]{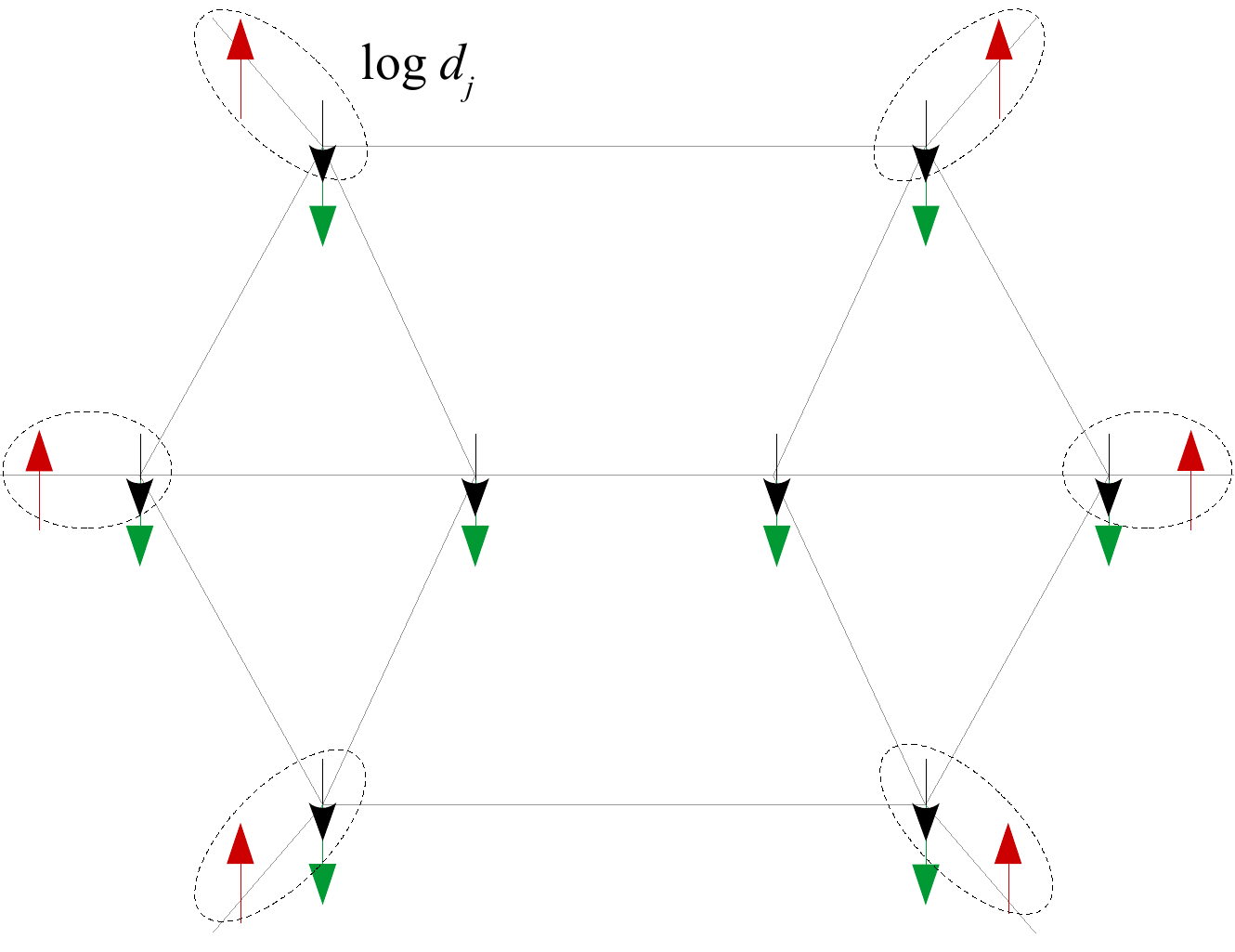}\caption{For $\log D_j \gg \log d_j$ the minimal energy configuration for $\mathcal{A}_1$ is that with all Ising spins $\sigma_v$ (shown in black) pointing down. Each pair of misaligned Ising- and boundary-spin carries a contribution to the free energy $F_1$ equal to $\log d_j$, so $F_1=|\partial \gamma| \log d_j$.}
		\label{rosso}
	\end{minipage}
\hspace{0.5cm}
	\begin{minipage}[t]{0.46\linewidth}
		\centering
		\includegraphics[width=0.9\linewidth]{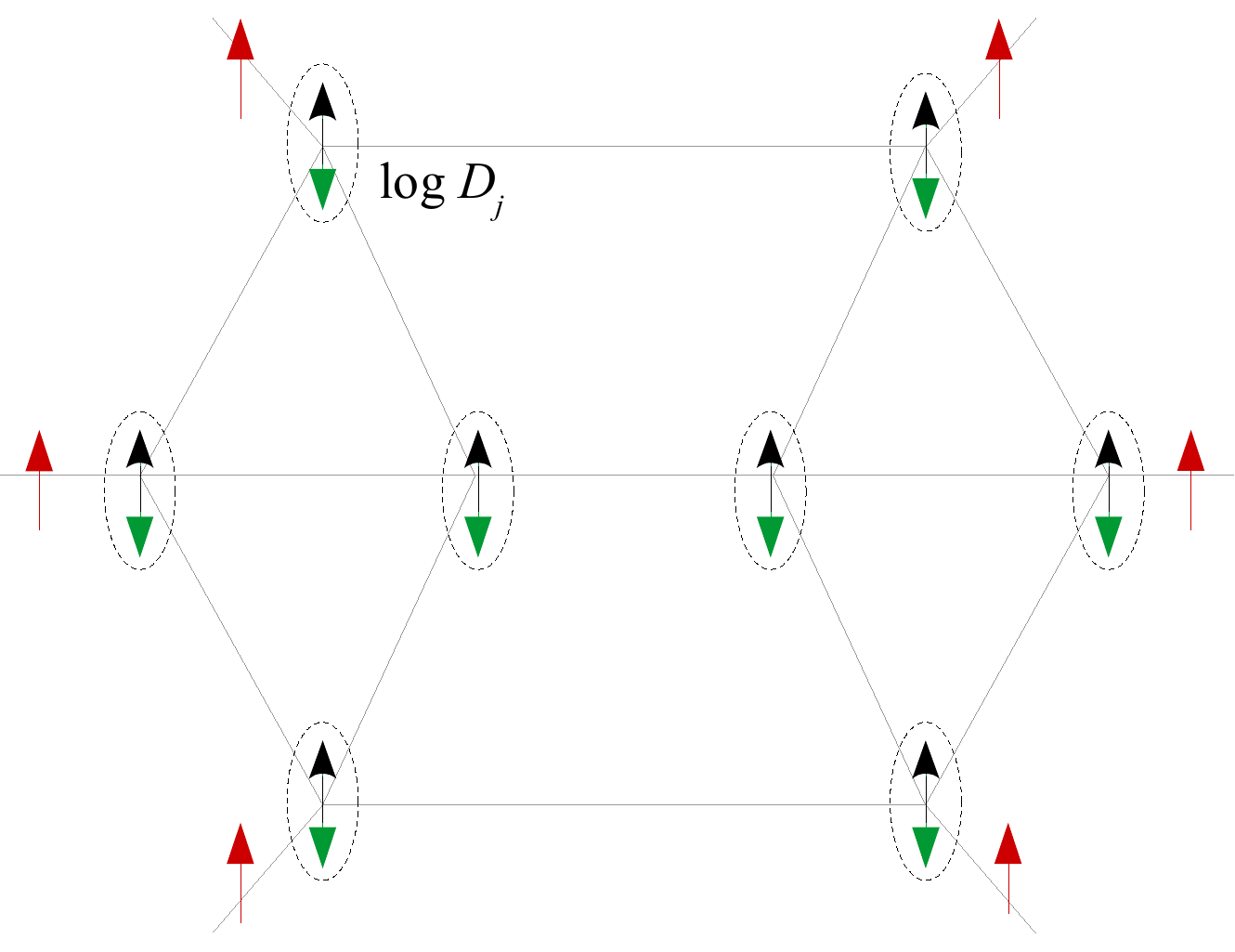}\caption{For $\log d_j \gg \log D_j$ the minimal energy configuration is that with all Ising spins $\sigma_v$ (shown in black) pointing up. Each pair of misaligned Ising- and bulk-spin carries a contribution to the free energy $F_1$ equal to $\log D_j$, so $F_1=N \log D_j$}
		\label{blu}
	\end{minipage}
\end{figure}

\begin{equation}\begin{split}\label{Ia}
\mathcal{A}(\vec{\sigma})=& -\frac{1}{2}\log d_j\left[\sum_{e_{vw}^i\in L}(\sigma_v\sigma_w -1)
 + \sum_{e_v^i\in \partial \gamma}(\sigma_v\mu_v-1)\right] -\frac{1}{2}\log D_j\sum_{v}(\sigma_v\nu_v-1) + k
\end{split}
\end{equation}
where we indicated as $D_j$ the dimension of the intertwiner recoupling $d$ $j$-spins (and omitted, for simplicity, the reference to the parametric dependence of the action from the pinning fields).

Let us start by evaluating the free energy $F_0$, which can be estimated (see \eq{approx}) by the minimum of $\mathcal{A}_0$, i.e. the Ising action with all pinning fields pointing up. It is easy to see that its minimum is reached when also the Ising spins point up, so that the interaction terms vanish, and the action coincides with the constant term. Since we are interested in a difference of free energies, the constant term can be set equal to zero; then $F_0\approx \mathcal{A}_{0\mathrm{min}}=0$.

The value of the free energy $F_1$, approximated by the minimum of $\mathcal{A}_1$ (Ising action in which the bulk fields are flipped down), depends instead on the relative strength of the interactions, specifically by the ratio of $\log d_j$ (interaction strength between Ising spins $\sigma_v$ and boundary spins $\mu_v$) to $\log D_j$ (interaction strength between Ising spins $\sigma_v$ and bulk spins $\nu_v$). 
Let us focus on the following two regimes, which are the counterpart of the asymptotic regimes considered in \cite{Hayden:2016cfa}: 
\begin{itemize}
	\item[-] $\log D_j \gg \log d_j$: the interaction with bulk spins is predominant, and the minimal energy configuration is thus the one with all Ising spins pointing down (see Figure \ref{rosso}). The contributions to the energy come from the misalignment of Ising spins with boundary spins, and amounts to $\log d_j$ for each boundary edge. The result is $F_1=|\partial \gamma| \log d_j$, where $|\partial \gamma|$ is the number of boundary edges. 
		\item[-] $\log d_j \gg \log D_j$: the interaction with boundary spins is predominant, and the minimal energy configuration is thus the one with all Ising spins pointing up (see Figure \ref{blu}). The contributions to the energy come from the misalignment of Ising spins with bulk spins, and amounts to $\log D_j$ for each vertex. The result is $F_1=N \log D_j$, which is the maximum possible value for the entropy of the reduced bulk state, as $(D_j)^N$ is the bulk-space dimension.
\end{itemize}
\begin{figure}[t]
	\centering
	\includegraphics[width=0.45\linewidth]{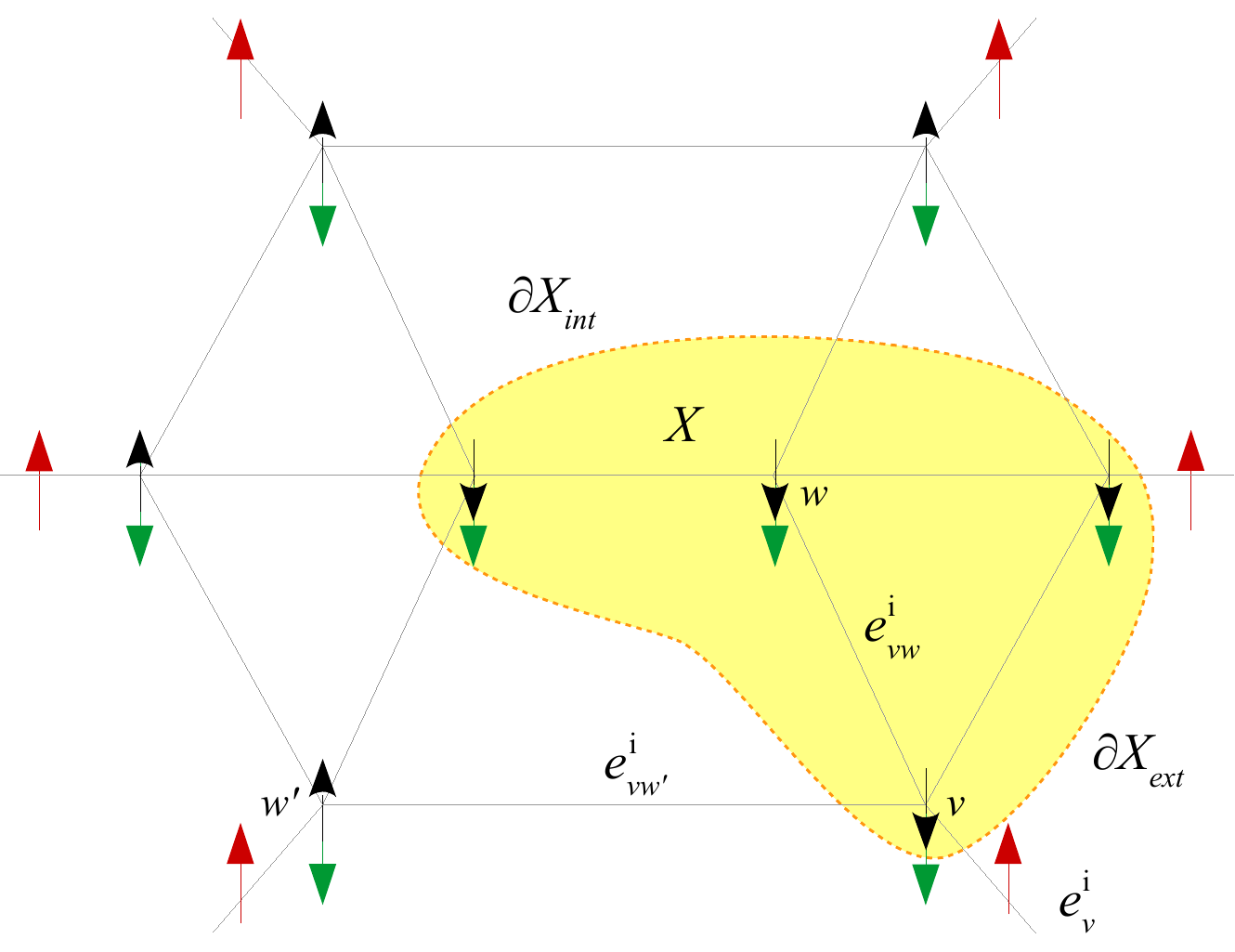}\caption{Ising configuration with Ising spins (shown in black) pointing down in a region $X$, bulk spins (in green) $\nu_v=-1$ $\forall v$ and boundary spins (in red) $\mu_v=+1$ $\forall v$. The contributions to $\mathcal{A}_1$ derive from: misaligned Ising- and boundary-spins ($\partial X_\text{ext}$), misaligned Ising- and bulk-spins (region outside $X$) and misaligned nearest-neighbor Ising spins ($\partial X_\text{int}$). }
	\label{regionX}
\end{figure}

We therefore found that, when the configuration that minimises $\mathcal{A}_1$ is the one with all Ising spins pointing up, the entropy reaches its maximum value, since $\overline{S_2(\rho_{\dot{\gamma}})}\approx F_1 -F_0 = N \log D_j$; this  corresponds to $\rho_{\dot{\gamma}}$ being completely mixed, namely to the bulk-to-boundary map being isometric. The isometry condition for the map can thus be translated to the requirement of stability of the all-up configuration: for every $\vec{\sigma}$ it must hold that $\mathcal{A}_1(\vec{\sigma})>N \log D_j$. Given an Ising configuration $\vec{\sigma}$, let $X(\vec{\sigma})$ be the region in which the Ising spins point down; by using \eq{Ia} the stability condition can then be written as
\begin{equation}\label{stab}
|\partial X(\vec{\sigma})| \log d_j >|X(\vec{\sigma})| \log D_j \quad \forall \vec{\sigma}
\end{equation}
where $|\partial X(\vec{\sigma})|=\# e \in \partial X(\vec{\sigma})$, and $|X(\vec{\sigma})|=\# v \in X(\vec{\sigma})$. 

The significant cases are the ones with vertex valence $d\ge 4$ (for $d\le3$ the intertwiner degree of freedom is suppressed). The dimension of the intertwiner space is given by \cite{Freidel:2010xna} \begin{equation}
D_j=\frac{2}{\pi}\int_{0}^{\pi}\diff \theta \sin^2\left(\frac{\theta}{2}\right)\left(\frac{\sin\left((j+\frac{1}{2})\theta\right)}{\sin\left(\frac{\theta}{2}\right)}\right)^d
\end{equation}
and the condition in \eq{stab} thus becomes the following:
\begin{equation}
e^{\frac{|\partial X(\vec{\sigma})|}{|X(\vec{\sigma})|}}  d_j > \frac{2}{\pi}\int_{0}^{\pi}\diff \theta \sin^2\left(\frac{\theta}{2}\right)\left(\frac{\sin\left(d_j \frac{\theta}{2}\right)}{\sin\left(\frac{\theta}{2}\right)}\right)^d.
\end{equation}
The interesting case for 4D quantum gravity models is $d=4$. In this case the dimension of the intertwiner space is simply given by $D_j = 2j+1=d_j$ \cite{Barbieri:1997ks}, i.e. it corresponds to the dimension of the edge spins, and the stability condition becomes
\begin{equation}\label{4}
|\partial X(\vec{\sigma})| >|X(\vec{\sigma})| \quad \forall \vec{\sigma}
\end{equation}
For the class of graphs we are considering (that is, at most one boundary link for each vertex) \eq{4} cannot be satisfied by the configuration with all Ising spins pointing down; the bulk-to-boundary map thus cannot be isometric.

\subsubsection{Inhomogeneous case}
We then consider the situation in which the spin assignment to the graph is not homogeneous.
Similarly to the previous case, the constant term in the Ising action can be set equal to zero, so that $F_0=0$ and the entropy can be estimated via $F_1$, and thus by the minimum of $\mathcal{A}_1$, with $k=0$. The key observation is, once again, the fact that when the all-up configuration is stable, namely minimises the action, $\overline{S_2(\rho_{\dot{\gamma}})}$ is maximised. In fact,  
\begin{equation}\begin{split}
 \mathcal{A}_1(\{\sigma_v=+1~ \forall v\})=\sum_{v}\log D_{\vec{j}_v} =\log D_{\dot{\gamma}},
\end{split}
\end{equation}
which is the maximum possible value for the entropy of the reduced bulk state. We thus need to require that, for any Ising configuration $\vec{\sigma}$, it holds that $\mathcal{A}_1(\vec{\sigma})>\log D_{\dot{\gamma}}$. By identifying a configuration $\vec{\sigma}$ through its spin-down region $X(\vec{\sigma})$ (also indicated just as $X$, when the explicit reference to the corresponding vector $\vec{\sigma}$ is unnecessary)  we can write   $\mathcal{A}_1(\vec{\sigma})$ as follows:
\begin{equation}\begin{split}
\mathcal{A}_1(\vec{\sigma})=&\sum_{e_{vw}^i\in \partial X_\text{int}}\log d_{j_{vw}^i}
+\sum_{e_v^i\in \partial X_\text{ext}}\log d_{j_v^i} +\sum_{v\in \overline{X}}\log D_{\vec{j}_v}
\end{split}
\end{equation}
where $\partial X_\text{int}=L \cap\partial X$ is the internal boundary of $X$, $\partial X_\text{ext}=\partial \gamma \cap \partial X$  the external one (see Figure \ref{regionX}) and $\overline{X}$ the complement of $X$; the inequality $\mathcal{A}_1(\vec{\sigma})>\log D_{\dot{\gamma}}$ then reads
\begin{equation}\begin{split}
&\sum_{e_{vw}^i\in \partial X_\text{int}}\log d_{j_{vw}^i}
+\sum_{e_v^i\in \partial X_\text{ext}}\log d_{j_v^i} >\sum_{v\in X}\log D_{\vec{j}_v},
\end{split}
\end{equation}
namely, in a more compact form,
\begin{equation}\begin{split}\label{dis}
	&\sum_{e_v^i \in \partial X}\log d_{j_v^i} >\sum_{v\in X}\log D_{\vec{j}_v},
	\end{split}
	\end{equation}
which implies
	\begin{equation}\begin{split}\label{diss}
	&\prod_{e_v^i \in \partial X}d_{j_v^i} >\prod_{v\in X} D_{\vec{j}_v} .
	\end{split}
	\end{equation} 
Therefore, for every spin-down region $X$ the dimension of the boundary must be greater then the dimension of the bulk.

Let us focus on the case of valence $d=4$. The dimension of the intertwiner space is given by the following expression~\cite{Barbieri:1997ks}:
\begin{equation}\small
D_{\vec{j}}	=\min\{j^1+j^2,j^3+j^4\}-\max\{|j^1-j^2|,|j^3-j^4|\}+1
\end{equation}
Note that when the spin assignment reduces to the homogeneous one, the dimension of the intertwiner space is $D_{\vec{j}}= 2j+1=d_j$ and we recover from Eq.~\eqref{diss} the inequality $|\partial X|>|X|$ found in the homogeneous case, which is clearly false.

Before moving to the general case, we consider the simplified scenario of graphs made of vertices with spins pairwise equal: $j_v^i \in \{j_v^\text{min},j_v^\text{max}\}$ $\forall  e_v^i \in \gamma$. For each vertex it thus holds that 
\begin{equation}\label{dm}
	D_{\vec{j}_v} = d_{j_v^\text{min}}.
	\end{equation}
	Combining Eq.~\eqref{dis} with Eq.~\eqref{dm} we then obtain
	\begin{equation}\begin{split}
	&\sum_{ \partial X}\log d_{j_v^i} > \sum_{v\in X}\log d_{j_v^\text{min}} ,
	\end{split}
	\end{equation}
	which is violated when, for example, the spin-down region $X$ is such that all its vertices have a link $e_v^i\in \partial X$ carrying the minimum spin $j_v^\text{min}$. Therefore, also in the case of vertices with spins pairwise equal the map is not an isometry.

We now consider the generic case. For vertex spins $j^\text{min}=j^a\le j^b \le j^c \le j^d=j^\text{max}$ the dimension of the intertwiner space is
\begin{equation}\label{di}
D_{j^aj^bj^cj^d} = \min\{j^a+j^d,j^b+j^c\}-\Delta+1
\end{equation}
where $\Delta =j^\text{max} - j^\text{min}$. 
 Combining Eq.~\eqref{di} with Eq.~\eqref{dis} we obtain the inequality
\begin{equation}\begin{split}\label{Delta}
	&\sum_{ e_v^i \in \partial X}\log d_{j_v^i} >\sum_{v\in X}\log \left( \min\{j_v^a+j_v^d,j_v^b+j_v^c\}-\Delta_v+1 \right)
	\end{split}
	\end{equation}
Let us stress that the key factors for the condition of \eq{Delta} to be satisfied are the following:
	\begin{itemize}
		\item[-] Combinatorial structure of the graph (number of links on the boundary $\partial X$ and number of vertices inside $\partial X$).
		\item[-] Dimensions (equivalently, spins) of the links; note that $\Delta$ \virg{measures} the difference with the homogeneous case, which corresponds to $\Delta=0$, and for which the isometry condition cannot be satisfied. 
\end{itemize}
	From Eq.~\eqref{Delta} we see that, for a given combinatorial structure, increasing $\Delta$, namely reducing the dimension of the bulk space, favours the attainment of the isometry condition. Since $j^d \le j^a+j^b+j^c$ \cite{Barbieri:1997ks}, it follows that $\Delta\le j^b + j^c$. For the maximum possible value $\Delta^\text{max} \coloneqq j^b+j^c$ the dimension of the intertwiner space is equal to 1 (since $\Delta=j^d-j^a=j^b+j^c$ implies $ j^a+j^d=j^b+j^c+2j^a$, therefore $\min\{j^a+j^d,j^b+j^c\}=j^b+j^c$), which means that the bulk degrees of freedom are suppressed; Eq.~\eqref{Delta} in fact becomes trivial:
			\begin{equation}\begin{split}
			&\sum_{ e_v^i \in \partial X}\log d_{j_v^i} >0.
			\end{split}
			\end{equation}
			The largest possible value of $\Delta$ which does not trivialize the bulk degrees of freedom, $\Delta=\Delta_{\text{max}}-1$, corresponds to an intertwiner space of dimension 2; when $\Delta=\Delta_{\text{max}}-1$ for all vertices the stability condition for the all-up configuration becomes
			\begin{equation}\begin{split}
			&\sum_{ e_v^i \in \partial X(\vec{\sigma})}\log d_{j_v^i} >|X(\vec{\sigma})| \log 2 \quad \forall \vec{\sigma},
			\end{split}
			\end{equation}
			which, for a given combinatorial pattern\footnote{We are considering graphs whose vertices can have at most one open edge, and in which two vertices cannot share more than one link.}, is satisfied for sufficiently high values of the edge spins.
		
		Let us compare the above analysis with that of Hayden et al. \cite{Hayden:2016cfa} for random tensor networks. 	
		In \cite{Hayden:2016cfa} the dimensions of links (\virg{bond dimensions} in the tensor network language) are equal; this corresponds to our homogeneous case. Moreover, in \cite{Hayden:2016cfa} the dimension of the bulk degrees of freedom is independent from the bond dimensions (in contrast to the intertwiner space, which depends on the spins attached to the vertex links) and can be chosen small enough to make the isometry condition satisfied. In our framework the link dimensions can differ, and the homogeneous configuration turns out to be the furthest from being isometric. This is a result of the correlation between bond dimensions and bulk degrees of freedom. In \cite{Hayden:2016cfa}, where such a correlation is absent, the homogeneous configuration can meet the isometry condition with a suitable choice of the bulk dimension.
		
		Let us finally stress the following points:
		\begin{itemize}
			\item[$\cdot$] When spins are assigned to the links randomly, $\Delta$ measures the spread of the corresponding probability distribution. A growing spread $\Delta$ corresponds to an increasingly uniform  probability distribution. This could be seen also as a condition on generic states in which spins are superposed, a class of which we are going to consider in the following.
			
			\item[$\cdot$] Increasing $\Delta$ increases the \virg{disorder} of the vertex structure, but decreases the dimension of the intertwiner space and thus the value of the maximum possible entropy of the bulk state.
			
			\item[$\cdot$] From the perspective of the effective Ising model, increasing $\Delta$ corresponds to reducing the minimum possible free energy.
		\end{itemize}

\section{Bulk/boundary entropy (fixed-spins case)}\label{entropy}

In this section we show how to compute, for a random graph state with fixed spins, the second order Rényi entropy of an arbitrary region of the graph, which may include part of the bulk (intertwiners attached to the vertices) and/or of the boundary (open edges of the graph), adapting the technique of \cite{Hayden:2016cfa} to our framework. We start by recalling that, when applying the replica trick, the second order Rényi entropy of a reduced state $\rho_P=\Tr_{\overline{P}}(\rho)$ is given by
\begin{equation}\label{trick}
S_2(\rho_P)=-\log \Tr\left(\rho_P^2 \right)=-\log \left(\frac{Z_1}{Z_0}\right)  
\end{equation}
with
\begin{equation}\begin{split}
\label{z_1}
&Z_1 \coloneqq \Tr \left[\left(\rho \otimes \rho\right)\mathcal{S}_P\right]\\&Z_0 \coloneqq \Tr \left[\rho \otimes \rho\right]
\end{split}
\end{equation}
where $\mathcal{S}_P$ is the swap operator acting on the two copies of subsystem $P$.

Consider the state $\ket{\phi_\gamma}=\left(\bigotimes_{e\in L} \bra{e}\right)\bigotimes_v \ket{f_v}$ with entanglement graph having connectivity  $\gamma$. 
We can write an arbitrary portion of the latter as $A\cup \Omega$, where $A$ is a region of the boundary $\partial \gamma$ (set of open edges), and $\Omega$ a region of the bulk $\dot{\gamma}$. Our goal is to compute the entropy of $A\cup \Omega$ for the state $\ket{\phi_\gamma}$. The density matrix of the latter can be written as
\begin{equation}\begin{split}
\rho=\ket{\phi_\gamma}\bra{\phi_\gamma}= \Tr_{L} \left(\bigotimes_{e \in L} \ket{e}\bra{e}\bigotimes_v \ket{f_v}\bra{f_v}\right),
\end{split}
\end{equation}
and the quantities $Z_1$ and $Z_0$ thus take the following form:
\begin{equation}\begin{split}\label{trace}
Z_{1/0} &= \Tr\left[\left(\rho \otimes \rho\right)\mathcal{S}_{A \cup \Omega}/\mathbb{I}\right]\\&=\Tr \left[\rho_L ^{\otimes 2}\bigotimes_v \left(\ket{f_v}\bra{f_v}\right)^{\otimes 2}\mathcal{S}_{A \cup \Omega}/\mathbb{I}\right]
\end{split}
\end{equation}
where $\rho_L \coloneqq \bigotimes_{e \in L} \ket{e}\bra{e}$, and the  trace is over all degrees of freedom, i.e. $ \Tr= \Tr_{\dot{\gamma}+L+\partial\gamma}$. 

We look at the case in which the vertex states are assigned randomly according to the uniform probability distribution. As we are going to show in the following, the key point of the  entropy calculation is that the randomization over the vertex states can be implemented \textit{before} performing the trace in \eq{trace}, due to the linearity of such operations.

\subsubsection{Random tensors and average entropy}\label{random}
As mentioned in Section \ref{average}, the tensor $(f_v)_{\vec{n} \iota}^{\vec{j}}$ associated to the vertex $v$ can be regarded as a vector of dimension $
D_v(\vec{j})\coloneqq d_{j^1}\cdot ... \cdot d_{j^d} \cdot  D_{\vec{j}}$. In analogy with \cite{Hayden:2016cfa}, to consider random tensors $(f_v)^{\vec{j}}_{\vec{n} \iota}$ with uniform probability distribution and compute the average of functions of them, we set $\ket{f_v}=U \ket{f^0_v}$, where $\ket{f^0_v}$ is a reference state and $U\in SU(D_v)$, and integrate over $U$ with the Haar measure. In the case of large spins we have that
\begin{equation}\label{limit}
\overline{S_2}\approx-\log \frac{\overline{Z_1}}{\overline{Z_0}},
\end{equation}
and the calculation of the average value of the entropy $S_2(\rho_{A\cup \Omega})$ thus reduces to the evaluation of the following quantities:
\begin{equation}\begin{split}\label{eqZ}
&\overline{Z_{1/0}}=  \Tr \left[\rho_L^{\otimes 2}\bigotimes_v \left(\overline{\ket{f_v}\bra{f_v}\otimes\ket{f_v}\bra{f_v}}\right)\mathcal{S}_{A \cup \Omega}/\mathbb{I}\right]
\end{split}
\end{equation}
where $\mathcal{S}_{A \cup \Omega}/\mathbb{I}$ refers to the presence of the swap operator $\mathcal{S}_{A \cup \Omega}$ for $\overline{Z_{1}}$ and of the identity $\mathbb{I}$ for $\overline{Z_{0}}$. 

As we show in Appendix \ref{schur}, from the Schur's lemma it follows that
\begin{equation}\label{eqschur}
\overline{\ket{f_v}\bra{f_v}\otimes\ket{f_v}\bra{f_v}} = \frac{\mathbb{I} + \mathcal{S}_v}{D_v^2 + D_v}
\end{equation}
where $\mathcal{S}_v$ is the operator swapping the two copies of the Hilbert space associated to vertex $v$ (i.e. to both edges and intertwiner degrees of freedom).
For $\overline{Z_{1/0}}$ we then obtain
\begin{equation}\begin{split}\label{z1}
&\overline{Z_{1/0}}=  \prod_v  \frac{1}{D_v^2 + D_v}\Tr \left[\rho_L^{\otimes 2}\bigotimes_v \left( \mathbb{I} + \mathcal{S}_v\right)\mathcal{S}_{A \cup \Omega}/\mathbb{I}\right].
\end{split}
\end{equation}
The r.h.s of Eq.~\eqref{z1} is a sum of $2^N$ terms with $\mathbb{I}$ or $\mathcal{S}_v$ for each vertex $v$. By introducing a two-level variable (an Ising spin, see below) $\sigma_v$ for each vertex $v$, whose value is associated to the presence of $\mathbb{I}$ ($\sigma_v=+1$) or $\mathcal{S}_v$ ($\sigma_v=-1$), this sum can be written as a partition function of the spins $\sigma_1, ...,\sigma_N$; in fact, each of the $2^N$ terms correspond to a given configuration $\vec{\sigma}\coloneqq \{\sigma_1, ...,\sigma_N\}$ of them. In particular,
\begin{equation}\begin{split}
\overline{Z_{1}}&=\sum_{\vec{\sigma}}  \prod_v  \frac{1}{D_v^2 + D_v}\Tr \left[\rho_L^{\otimes 2}\mathcal{S}_{A \cup \Omega}\prod_{v:\sigma_v=-1} \mathcal{S}_v\right]=\sum_{\vec{\sigma}}e^{-\mathcal{A}_1(\vec{\sigma})},
\end{split}
\end{equation}
where
\begin{equation}\begin{split}\label{a11}\small
\mathcal{A}_1(\vec{\sigma})=& -\log  \left(\prod_v  \frac{1}{D_v^2 + D_v}\right) -\log  \left(\Tr \left[\rho_L^{\otimes 2}\left(\prod_{v:\sigma_v=-1} \mathcal{S}_v\right)\mathcal{S}_{A \cup \Omega}\right]\right)
\end{split}
\end{equation}
and similarly for $\overline{Z_{0}}$, whose action $\mathcal{A}_0(\vec{\sigma})$ has the identity operator $\mathbb{I}$ in place of $\mathcal{S}_{A \cup \Omega}$.
Since the swap operator $\mathcal{S}_v$ acts independently on each degree of freedom of vertex $v$, it factorizes as follows:
\begin{equation}
\mathcal{S}_v=\prod_{i=0}^d\mathcal{S}^i_v
\end{equation}
where $\mathcal{S}^i_v$, with $i=1,...,d$, is the operator swapping the two copies of the Hilbert space associated to the edge of colour $i$, while $\mathcal{S}^0_v$ is the operator swapping the two copies of the intertwiner space. Thanks to this factorization, the trace of Eq.~\eqref{a11} can be split into the contributions coming from the bulk ($\dot{\gamma}$), the internal links ($L$) and the boundary links ($\partial \gamma$):
\begin{equation}\begin{split}\label{tracee}
&\Tr \left[ \rho_L^{\otimes 2}\left(\prod_{v:\sigma_v=-1} \mathcal{S}_v\right)\mathcal{S}_{A\cup \Omega}\right]= \Tr_{\dot{\gamma}} \left[\mathcal{S}_{\Omega}\prod_{v: \sigma_v=-1} \mathcal{S}^0_v\right]\Tr_{{L}} \left[ \rho_L^{\otimes 2}\prod_{e_v^i\in L: \sigma_v=-1} \mathcal{S}^i_v\right]
\Tr_{\partial\gamma}\left[\mathcal{S}_{A}\prod_{e_v^i\in \partial\gamma: \sigma_v=-1} \mathcal{S}^i_v\right]
\end{split}
\end{equation}
Then, for every boundary link $e_v^i$ (coming out of a vertex $v$) we can introduce a spin $\mu_v=\pm 1$ which determines the presence ($\mu_v=-1$) or absence ($\mu_v=+1$) of an additional swap operator $\mathcal{S}^i_v$ coming from $\mathcal{S}_A$, depending on whether or not $e_v^i$ belongs to $A$. Similarly, for each vertex $v$ we introduce a spin $\nu_v \pm 1$ which signals whether or not it belongs to $\Omega$, i.e. the presence ($\nu_v=-1$) or absence ($\nu_v=+1$) of an additional swap operator $\mathcal{S}^0_v$ coming from $\mathcal{S}_\Omega$. Note that for  $\mathcal{A}_0(\vec{\sigma})$, which does not contain the operator $\mathcal{S}_{A \cup \Omega}$, $\mu_v=\nu_v=+1$ for all vertices. 

We finally obtain the action of a classical Ising model defined on the graph $\gamma$, with additional \virg{pinning fields} \cite{Hayden:2016cfa} for the bulk ($\vec{\nu}\coloneqq\{\nu_v\}$) and boundary ($\vec{\mu}\coloneqq\{\mu_v\}$) degrees of freedom (see Appendix \ref{det} for details on the derivation of the various terms):
\begin{equation}\begin{split}\label{action}
\mathcal{A}_{1}(\vec{\sigma})=&\sum_v \log \left(D_v^2 + D_v\right)  -\frac{1}{2}\left[ \sum_{e_{vw}^i\in L}(\sigma_v\sigma_w -1)\log d_{j_{vw}^i}
+\sum_{e_v^i\in \partial \gamma}(3+\sigma_v\mu_v)\log d_{j_v^i} +\sum_{v}(3+\sigma_v\nu_v)\log D_{\vec{j}_v}\right]\end{split}
\end{equation}
The first term on the r.h.s. of \eq{action} can be decomposed into edge and intertwiner dimensions as follows:
\begin{equation}\begin{split}
\sum_v \log \left(D_v^2 + D_v\right)  =& 2\sum_v \log D_v + \sum_v \log \left(1 + D_v^{-1}\right)=2 \sum_{e_v^i\in \gamma} \log d_{j_v^i} + 2 \sum_v \log D_{\vec{j}_v}+ \sum_v \log \left(1 + D_v^{-1}\right)
\end{split}
\end{equation}
and \eq{action} then becomes
\begin{equation}\begin{split}\label{action2}
\mathcal{A}_{1}(\vec{\sigma})=& -\frac{1}{2} \left[\sum_{e_{vw}^i\in L}(\sigma_v\sigma_w -1)\log d_{j_{vw}^i}
+\sum_{e_v^i\in \partial \gamma}(\sigma_v\mu_v-1)\log d_{j_v^i} +\sum_{v}(\sigma_v\nu_v-1)\log D_{\vec{j}_v}  \right]+ k
\end{split}
\end{equation}
where $k$ is a constant contribution given by
\begin{equation}
    k=2 \sum_{e_v^i\in \gamma/\partial \gamma} \log d_{j_v^i} +  \sum_v \log \left(1 + D_v^{-1}\right).
\end{equation}
The action $\mathcal{A}_0(\vec{\sigma})$takes the same form but, as mentioned before, with all pinning fields equal to $+1$.

By defining the free energies $F_{1/0}=-\log \overline{Z_{1/0}}$, the expression of \eq{limit} for the average entropy in the high spins regime becomes 
\begin{equation}\label{free}
\overline{S_2(\rho_{A\cup \Omega})}\approx F_1 - F_0.
\end{equation}
Moreover, in this regime (corresponding, for the classical Ising model, to the low temperature regime) the partition functions $\overline{Z_{1/0}}$ can be estimated by the lowest energy configuration, therefore $F =-\log \overline{Z} \approx \mathcal{A}_{\text{min}}$. Since we are interested only in the difference of free energies, we can set $k=0$. Note also that, for the action $\mathcal{A}_0(\vec{\sigma})$ (in which all pining fields are equal to $+1$), the minimal energy configuration is the one with all Ising spins pointing up: $\sigma_v=+1$ $\forall v$; when setting $k=0$, we have that $F_0 \approx \mathcal{A}_{0\text{min}}=0$. The computation of $\overline{S_2(\rho_{A\cup \Omega})}$ then reduces to that of $F_1\approx \mathcal{A}_{1\text{min}}$.

\section{Isometry condition for bulk-to-boundary maps (general case with spin superposition)}\label{isosum}
In this section we focus on the isometry condition for the bulk-to-boundary maps introduced in Section \ref{genericmap}, defined by graph states constructed out of the gluing of vertex wavefunctions $f_v$ that spread over all possible values of the edge spins, i.e.
\begin{equation}\label{ver}
\ket{f_v} =\bigoplus_{\vec{j}}\sum_{\vec{n}\iota}(f_v)^{\vec{j}}_{\vec{n} \iota} \ket{\vec{j}\vec{n}\iota}.
\end{equation}
We recall that an example of these states is given by
\begin{equation}\begin{split}\label{ex}
\ket{{\phi_\gamma}}&=\bigotimes_{e\in \gamma} \bra{e}\bigotimes_v \ket{f_v}
\end{split}
\end{equation}
where $\ket{e}$ are maximally entangled states of edges glued together to form internal links of the graph $\gamma$, including all possible values of edge spins (see \eq{edge}). In terms of the components of the corresponding bulk-to-boundary map $M$, the graph state $\ket{{\phi_\gamma}}$ can be written as
\begin{equation}\begin{split}
\ket{{\phi_\gamma}}=\bigoplus_{J_{\partial}}\bigoplus_{J|_{J_{\partial}}}\sum_{\alpha_J} \left(M \otimes \mathbb{I}\right) \ket{\alpha_J}\otimes \ket{\alpha_J}
\end{split}
\end{equation}
where $\ket{\alpha_J}$ is a basis for the bulk Hilbert space having spins $J$, $M_J$ is the map restricted to the $J$-subspace, and the notation $J|_{J_{\partial}}$ means that the subset of $J$ relative to the boundary spins has been fixed to $J_{\partial}$.
The reduced bulk state can thus be expressed as follows:
\begin{equation}\begin{split}
\Tr_{\partial\gamma}\left(\ket{{\phi_\gamma}}\bra{{\phi_\gamma}}\right)= \bigoplus_{J|_{J_{\partial}}}\bigoplus_{J'|_{J_{\partial}}} \left(M^{\dagger}M \right)_{\alpha'_{J'}\alpha_J} \ket{\alpha_J}   \bra{\alpha'_{J'}}.
\end{split}
\end{equation}
Therefore, analogously to what happens in the fixed-spins case, the isometry condition $\left(M^{\dagger}M  \right)_{\alpha'_{J'}\alpha_J} = \delta_{\alpha'_{J'}\alpha_J}$ translates to the requirement that the reduced bulk state is proportional to the identity, i.e. is a maximally mixed state, a requirement that can be checked through an entropy calculation. We focus again on the second-order Rényi entropy computed through the replica trick according to \eq{trick} and \eq{z_1}, and assume a uniform probability distribution for the random vertex wavefunctions $f_v$; we also introduce a lower bound $j_\text{min}$ for the edge spins large enough to ensure that
\begin{equation}\label{s2}
\overline{S_2(\rho_{\dot{\gamma}})}\approx-\log \frac{\overline{Z_1}}{\overline{Z_0}}.
\end{equation}
As a first step, we implement the randomization over the vertex wavefunctions of \eq{ver}. In particular, randomize each $\vec{j}$-sector separately, by rotating every $(f_v)^{\vec{j}}_{\vec{n} \iota}$ component via $U_{\vec{j}}\in SU(D_v(\vec{j}))$ and integrating over the latter with the Haar measure, analogously to what has been done in the fixed-spins case. For the Schur's lemma, the randomization procedure then yields
\begin{equation}\begin{split}
\overline{\ket{f_v}\bra{f_v}\otimes\ket{f_v}\bra{f_v}}= \bigoplus_{\vec{j}}\frac{\mathbb{I} + \mathcal{S}_v^{[\vec{j}]}}{D_v(\vec{j}) \left( D_v(\vec{j})+1\right)}.
\end{split}
\end{equation}
where $[\vec{j}]$ refers to the $\vec{j}$-spin subspace on which the identity and swap operators act.
The quantities $\overline{Z_1}$ and $\overline{Z_0}$ can then be expressed as Ising partition functions with the following action  (see Appendix \ref{sumspins} for details):
\begin{equation}\begin{split}\label{ag}
\mathcal{A}(\vec{\sigma})=&-\log \mathcal{C}  -\sum_v\log \left(\sum_{\vec{j}_v}D_{\vec{j}_{v}}^{\frac{1}{2}(3 + \sigma_v\nu_v)}\right)-\sum_{e_{vw}^i\in L}\log 
\left(\sum_{j_{vw}^i}~d_{j_{vw}^i}^{\frac{1}{2}(\sigma_v\sigma_w -1)}\right)
-\sum_{e_v^i\in \partial \gamma}\log \left( \sum_{j_v^i}~ d_{j_v^i}^{\frac{1}{2}(3+\sigma_v\mu_v)}\right)
\end{split}
\end{equation}
with
\begin{equation}
\mathcal{C}\coloneqq \prod_v \sum_{\vec{j}_v}\frac{1}{D_v(\vec{j}) \left( D_v(\vec{j})+1\right)}
\end{equation}
and where an upper bound $j_\text{max}$ on the spin values has been introduced to regularise the calculation. In the large spins regime (i.e. large lower bound $j_\text{min}$) the partition function is dominated by the lowest energy configuration, and the average entropy becomes $
\overline{S_2(\rho_{\dot{\gamma}})}\approx \mathcal{A}_{1\text{min}} - \mathcal{A}_{0\text{min}}$. 
Therefore, the bulk-to-boundary map is an isometry if the latter reaches the value
\begin{equation}\begin{split}\label{max}
\overline{S_2(\rho_{\dot{\gamma}})}_\text{max}=\sum_{v=1}^N\log 
\left(\sum_{\vec{j}_{v}}~D_{\vec{j}_{v}}\right)
\end{split}
\end{equation}
As shown in Appendix \ref{sumspins}, the action $\mathcal{A}_0(\vec{\sigma})$ is minimized by the all-up configuration $\sigma_v=+1$ $\forall v$. The minimization of $\mathcal{A}_1(\vec{\sigma})$ is instead related to the combinatorial structure of the graph $\gamma$; when almost every vertex has an edge on the boundary, i.e. $|\partial \gamma|\approx N$ (where $|\partial \gamma|$ is the number of edges in $\partial \gamma$) the boundary-edge sum dominates, and $\mathcal{A}_1(\vec{\sigma})$ reaches the minimum value when all Ising spins point up. In this case the entropy is given by 
\begin{equation}\begin{split}
\overline{S_2(\rho_{\dot{\gamma}})}=\sum_v\log \frac{\left(\sum_{\vec{j}_v}D^2_{\vec{j}_{v}}\right)}{\left(\sum_{\vec{j}_v}D_{\vec{j}_{v}}\right)}
\end{split}
\end{equation}
which approaches the value of \eq{max} when $\sum_{\vec{j}_v}D^2_{\vec{j}_{v}}\approx \left(\sum_{\vec{j}_v}D_{\vec{j}_{v}}\right)^2$ (we recall that every sum is bounded by the values $j_\text{min}$ and $j_\text{max}$ for the edge spins).
In the case $|\partial \gamma|\ll N$ is instead the intertwiner sum to dominate, and $\mathcal{A}_1(\vec{\sigma})$ is minimized by the all-down configuration $\sigma_v=-1$ $\forall v$; then 
\begin{equation}\begin{split}
\overline{S_2(\rho_{\dot{\gamma}})}\approx \sum_{e_v^i\in \partial \gamma}\log \frac{\left( \sum_{j_v^i}~ d^2_{j_v^i}\right)}{\left( \sum_{j_v^i}~ d_{j_v^i}\right)}
\end{split}
\end{equation}
which, due to the assumption $|\partial \gamma|\ll N$, can reach the maximum entropy value given by \eq{max} only for sufficiently high values of the edge spins. 


\section{Conclusions and outlook}
We have defined bulk/boundary maps corresponding to quantum gravity states in the tensorial group field theory formalism, for quantum geometric models sharing the same type of quantum states of canonical loop quantum gravity and spin foam models; the definition of the maps relies on a partition of the quantum degrees of freedom associated to an open graph into bulk and boundary ones. We then identified the entanglement properties that a state must satisfy in order to the corresponding bulk-to-boundary map be an isometry. This isometry condition is necessary for an holographic behaviour of the quantum state. We have analysed different types of quantum states, identifying those that define isometric bulk/boundary maps, generalising the random tensor network techniques introduced in \cite{Hayden:2016cfa} to our tensorial spin networks.

Our results can now be developed in several directions.

First, remaining at the kinematical level, it is interesting to consider even more general quantum states involving sums over spin labels, which are not given by tensor networks themselves, even if this may require abandoning or at least modifying the use of random tensor network techniques we used in our work. Next, one further class of quantum states can be analysed, i.e. that involving superpositions of graphs themselves, in addition to superpositions of algebraic data. Not only these states represent the most general ones that appear in the Hilbert space of the theory, but superpositions of graphs are naturally produced by the quantum dynamics and thus we should expect that only holographic maps including such superpositions can be realized at the dynamical level. While the spaces of quantum states for fixed graph in group field theory and canonical loop quantum gravity coincide (for appropriately chosen GFT models), their full Hilbert spaces including graph superpositions are different. This means that the most general class of bulk/boundary maps will also require context-specific analysis, and have context specific properties, possibly highlighting interesting differences between these frameworks from the holographic perspective. 

Summing over graph structure is most likely an ingredient of any coarse-graining procedure for the quantum gravity states we use to define bulk/boundary maps. In turn, some such coarse-graining procedure may be needed to achieve an effective isometry property for the map, and holographic behaviour. One strategy would be that appropriate superpositions of the fundamental quantum gravity states with basic GFT tensors associated to the vertices could be replaced, after coarse-graining, by an effective state based on some \lq collective\rq graph with absolute maximally entangled (AME) (perfect) tensors \cite{PhysRevA.86.052335} associated to its vertices;for such effective entanglement graph state, the isometry condition would follow directly from the properties of the tensors associated to its vertices. 

In the GFT context, the bulk states live in a Fock space; it would be interesting to determine the conditions under which the bulk/boundary correspondence they define maps them into boundary data forming as well a Fock space. When this happens, we could expect that there is some GFT for boundary data only, and some hidden duality between the bulk and boundary GFTs. Of course, the duality should hold at the dynamical level to be truly interesting.

The contribution from the quantum dynamics is the important missing ingredient in our analysis. Imposition of the quantum dynamics selects first of all the physical quantum states of the theory, and it would be only these physical states that one should consider, in principle, to define holographic maps. Moreover, the quantum dynamics would contribute additional weights to the maps between quantum states we consider, and thus affect their properties and their (approximate) holographic behaviour. Last, imposition of the quantum dynamics provides an additional form of randomization for the vertex tensors, which needs also to be taken into account in an analysis along the lines we followed.
From a more physical perspective, our bulk/boundary maps need to be extended to include quantum geometric observables depending on bulk or boundary data, to explore which quantum states allow to map bulk observables to boundary ones (possibly in a dynamical context), for which observables and under which approximations.

Finally, the same techniques we used in this work allow to derive Ryu-Takanyagi entropy relations for our quantum gravity states; these relations, combined with the holographic maps we defined, can then be used to identify under which conditions the bulk entanglement ends up producing a region in the quantum states that behaves like a black hole, analogously to what has been found in \cite{Hayden:2016cfa}. This would be important because it may give important insights on the type of microstates corresponding to black holes in the GFT (and maybe canonical LQG) context. In the same direction, our methods and results should be applied to the class of candidate black hole microstates proposed in \cite{PhysRevD.97.066017,PhysRevLett.116.211301} and that were found to possess interesting holographic properties as well, in order to test the results obtained in that context from a different perspective.

\onecolumngrid
\appendix

\section{Calculations for the average entropy}\label{det}
In this section we present some details on the calculation of the average value of the entropy $S_2(\rho_{A\cup \Omega})$ presented in Section \ref{entropy}. In particular, we outline the proof of \eq{eqschur} and illustrate the computation of the internal-link, boundary-edge and bulk contributions deriving from the trace in \eq{tracee}.

\subsection{Randomization over (double copy of) vertex states}\label{schur}
A detailed proof of the result given by \eq{eqschur} can be found in \cite{2013arXiv1308.6595H};  we sketch here the same argument, adapted to our framework. The first step is to recognize that the space $\h_D \otimes_\mathrm{sym} \h_D$, where $\h_D$ is a $D$-dimensional Hilbert space and $\otimes_\mathrm{sym}$ is the symmetric tensor product, carries an irreducible representation of the group of $D\times D$ unitary matrices, under the map associating to any such matrix $U$ the double copy $U^{\otimes 2}$. Then, given $\ket{f}\in \h_D$, consider the density matrix 
\begin{equation}
	\rho \coloneqq \mathbb{E}_f \left[\ket{f}\bra{f} \otimes \ket{f}\bra{f}\right],
\end{equation}
where $\mathbb{E}_f[\cdot]$ is the average over $f$ according to an arbitrary probability distribution. Since $\rho$ commutes with all $U^{\otimes 2}$, by Schur's lemma (see e.g. \cite{tinkham2003group} for a formulation of the latter which fits the present argument) it must be proportional to the identity operator on $\h_D \otimes_\mathrm{sym} \h_D$, which is given by 
\begin{equation}
\frac{1}{2}\sum_{\pi \in S_2} P(\pi) =\frac{ \mathbb{I} + \mathcal{S}}{2},
\end{equation}
where $S_2$ is the symmetric group on 2 objects and $P(\pi)$ is the operator permuting vectors in $\h_D^{\otimes 2}$ according to the permutation $\pi \in S_2$, which is $\mathbb{I}$ for the trivial one (identity), and $\mathcal{S}$ for the swapping. 

\subsection{Internal-links contribution }\label{see}
The trace
\begin{equation}\begin{split}
\Tr_{{L}} \left[\rho_L^{\otimes 2}\prod_{e_v^i\in L: \sigma_v=-1} \mathcal{S}^i_v\right] = \Tr_{{L}} \left[ \left(\bigotimes_{e_{vw}^i\in L}\ket{e_{vw}^i}\bra{e_{vw}^i}\right)^{\otimes 2}\prod_{e_v^i\in L: \sigma_v=-1} \mathcal{S}^i_v\right]
\end{split}
\end{equation}
factorizes over single-link contributions. There are three different possibilities for the term related to a generic link $e_{vw}^i$, since it  can contain 

\begin{itemize}
	\item[1)] no swap operators ($\sigma_v=\sigma_w=+1$):
	\begin{equation}\begin{split}
	\Tr \left[\ket{e_{vw}^i}\bra{e_{vw}^i}\otimes \ket{e_{vw}^i}\bra{e_{vw}^i} \right]=& \Tr \left[  \frac{1}{d_j^2}\sum_{mnpq} \ket{m}_v\ket{m}_w \bra{n}_v\bra{n}_w \otimes \ket{p}_v\ket{p}_w \bra{q}_v\bra{q}_w\right]\\=&  \frac{1}{d_j^2}\sum_{mnpq} \delta_{mn}  \delta_{pq}=\frac{1}{d_j^2}\sum_{m} \delta_{mm}\sum_{p} \delta_{pp}=1;
	\end{split}
	\end{equation}
	\item[2)] just one swap operator ($\mathcal{S}_v^i/\mathcal{S}_w^i$ for $\sigma_v=-\sigma_w=-1/+1$):
	\begin{equation}\begin{split}
	\Tr \left[\ket{e_{vw}^i}\bra{e_{vw}^i}\otimes \ket{e_{vw}^i}\bra{e_{vw}^i} \mathcal{S}^i_v\right]=& \Tr \left[  \frac{1}{d_j^2}\sum_{mnpq} \ket{m}_v\ket{m}_w \bra{n}_v\bra{n}_w \otimes \ket{p}_v\ket{p}_w \bra{q}_v\bra{q}_w\mathcal{S}^i_v\right]\\=&\Tr \left[  \frac{1}{d_j^2}\sum_{mnpq} \ket{p}_v\ket{m}_w \bra{n}_v\bra{n}_w \otimes \ket{m}_v\ket{p}_w \bra{q}_v\bra{q}_w\right]\\=&  \frac{1}{d_j^2}\sum_{mnpq} \delta_{np} \delta_{mn} \delta_{mq}\delta_{pq}=\frac{1}{d_j^2}\sum_{m} \delta_{mm}=\frac{1}{d_j};
	\end{split}
	\end{equation}

	\item[3)] the two swap operators $\mathcal{S}_v^i$ and $\mathcal{S}_w^i$ ($\sigma_v=\sigma_w=-1$):
	\begin{equation}\begin{split}
	\Tr \left[\ket{e_{vw}^i}\bra{e_{vw}^i}\otimes \ket{e_{vw}^i}\bra{e_{vw}^i} \mathcal{S}^i_v\mathcal{S}^i_w\right]=& \Tr \left[  \frac{1}{d_j^2}\sum_{mnpq} \ket{m}_v\ket{m}_w \bra{n}_v\bra{n}_w \otimes \ket{p}_v\ket{p}_w \bra{q}_v\bra{q}_w\mathcal{S}^i_v \mathcal{S}^i_w\right]\\=&\Tr \left[  \frac{1}{d_j^2}\sum_{mnpq} \ket{p}_v\ket{p}_w \bra{n}_v\bra{n}_w \otimes \ket{m}_v\ket{m}_w \bra{q}_v\bra{q}_w\right]\\=&  \frac{1}{d_j^2}\sum_{mnpq} \delta_{np}  \delta_{mq}=\frac{1}{d_j^2}\sum_{n} \delta_{nn}\sum_{m} \delta_{mm}=1.
	\end{split}
	\end{equation}
\end{itemize}
For both cases $1)$ and $3)$ the link contribution to the average entropy is zero, as 
\begin{equation}
-\log \left(\Tr \left[\ket{e_{vw}^i}\bra{e_{vw}^i}\otimes \ket{e_{vw}^i}\bra{e_{vw}^i} \right]\right) =-\log \left(\Tr \left[\ket{e_{vw}^i}\bra{e_{vw}^i}\otimes \ket{e_{vw}^i}\bra{e_{vw}^i} \mathcal{S}^i_v\mathcal{S}^i_w\right]\right) = 0.
\end{equation}
In the case $2)$, the link $e_{vw}^i$ contributes to the average entropy with the quantity
	\begin{equation}
-\log \left(\Tr \left[\ket{e_{vw}^i}\bra{e_{vw}^i}\otimes \ket{e_{vw}^i}\bra{e_{vw}^i} \mathcal{S}^i_v\right]\right) = \log d_j.
\end{equation}

\subsection{Boundary-edges contribution}
The trace
\begin{equation}\begin{split}
\Tr_{\partial\gamma}\left[\mathcal{S}_{A}\prod_{e_v^i\in \partial\gamma: \sigma_v=-1} \mathcal{S}^i_v\right]
\end{split}
\end{equation}
factorizes over contributions coming from single boundary-edges. The trace term for an edge $e_v^i$ can contain
\begin{itemize}
\item[1)] no swap operators ($\sigma_v=\mu_v=1$) or two times the swap operator $\mathcal{S}^i_v$ ($\sigma_v=\mu_v=-1$); in both cases the operator inside the trace is the identity: 
\begin{equation}\begin{split}
\Tr\left[\left( \mathcal{S}^i_v\right)^2\right]=& \Tr \left[\mathbb{I}\otimes \mathbb{I} \right] =d_j^2
\end{split}
\end{equation}
    \item[2)] only one swap operator ($\sigma_v=-\mu_v$):
\begin{equation}\begin{split}
 \Tr\left[ \mathcal{S}^i_v\right]=& \Tr \left[ \sum_{mn} \ket{m}_v\bra{m}_v  \otimes\ket{n}_v\bra{n}_v  \mathcal{S}^i_v \right]=\Tr \left[ \sum_{mn} \ket{n}_v\bra{m}_v  \otimes\ket{m}_v\bra{n}_v   \right]= \sum_{m}\delta_{mm} =d_j 
\end{split}
\end{equation}
\end{itemize}
Case $1)$ is identified by the condition $\sigma_v\mu_v=1$, while case $2)$ by $\sigma_v\mu_v=-1$; the contribution of the boundary edges $e_v^i\in \partial \gamma$ to the average entropy can thus be expressed as follows:
\begin{equation}\begin{split}
-\log \Tr_{\partial\gamma}\left[\mathcal{S}_{A}\prod_{e_v^i\in \partial\gamma: \sigma_v=-1} \mathcal{S}^i_v\right]= -\sum_{e_v^i\in \partial \gamma}\frac{1}{2}(3+\sigma_v\mu_v)\log d_{j_v^i}.
\end{split}
\end{equation}
\subsection{Bulk contribution}
The trace
\begin{equation}\begin{split}
 \Tr_{\dot{\gamma}} \left[\mathcal{S}_{\Omega}\prod_{v: \sigma_v=-1} \mathcal{S}^0_v\right]
\end{split}
\end{equation}
factorizes over the intertwiners associated to the various vertices; the computation of the contribution for each vertex $v$ is analogous to the one presented in the previous section for boundary edges, the only difference being the dimension of the Hilbert space under consideration, i.e. $D_{\vec{j}_v}$ instead of $d_{j_v^i}$. The bulk contribution to the entropy thus takes the following form:
\begin{equation}\begin{split}
-\log  \Tr_{\dot{\gamma}} \left[\mathcal{S}_{\Omega}\prod_{v: \sigma_v=-1} \mathcal{S}^0_v\right]= -\sum_{v}\frac{1}{2}(3+\sigma_v\nu_v)\log D_{\vec{j}_v}.
\end{split}
\end{equation}
\section{Bulk/boundary average entropy for the general case (sum over spins)}\label{sumspins}
The quantities $\overline{Z_{1/0}}$ for the computation of the average second-order Rényi entropy of a generic region of the graph, composed of a portion $A$ of the boundary and $\Omega$ of the bulk, can be written as follows:
\begin{equation}\begin{split}
\overline{Z_{1/0}}=&  \Tr \left[\rho_L^{\otimes 2}\bigotimes_v \overline{\ket{f_v}\bra{f_v}\otimes\ket{f_v}\bra{f_v}}\mathcal{S}_{A\cup \Omega}/\mathbb{I}\right]\\=& \Tr \left[ \left(\bigoplus_{J_L}\left(\rho^{J_L}_{L}\right)^{\otimes 2}\right)\left(\bigotimes_v \bigoplus_{\vec{j}_v}\frac{\mathbb{I} + \mathcal{S}_v^{[\vec{j}_v]}}{D_v(\vec{j}_v) \left( D_v(\vec{j}_v)+1\right)}\right)\bigoplus_{J}\mathcal{S}^{[J]}_{\Omega}/\mathbb{I}\bigoplus_{J_\partial}\mathcal{S}^{[J_\partial]}_{A}/\mathbb{I}\right].
\end{split}
\end{equation}
where $J_L \coloneqq \{j_{vw}^i|e_{vw}^i \in L\}$  and
\begin{equation}
    \rho^{J_L}_L \coloneqq \bigotimes_{e\in L} \ket{e_{vw}^i}\bra{e_{vw}^i} \quad \mathrm{with}\quad \ket{e_{vw}^i}=\frac{1}{\sqrt{d_{j_{vw}^i}}}\sum_n  \ket{j_{vw}^i n }\otimes \ket{j_{vw}^in}
\end{equation}
Note that the non-diagonal (in the edge spins) components of $\rho_L$ do not appear because of the trace. 
By introducing Ising spins $\vec{\sigma}$ analogous to the ones considered in Section \ref{random}, the previous expression becomes
\begin{equation}\begin{split}\label{big}
\overline{Z_{1/0}}= \sum_{\vec{\sigma}} \mathcal{C} ~\Tr \left[\left(\bigoplus_{J_L}\left(\rho^{J_L}_{L}\right)^{\otimes 2}\right)\left(\bigotimes_{v:\sigma_v=-1} \bigoplus_{\vec{j}_v} \mathcal{S}_v^{[\vec{j}_v]}\right)\bigoplus_{J}\mathcal{S}^{[J]}_{\Omega}/\mathbb{I}\bigoplus_{J_\partial}\mathcal{S}^{[J_\partial]}_{A}/\mathbb{I}\right]
\end{split}
\end{equation}
where
\begin{equation}
\mathcal{C}\coloneqq \prod_v \sum_{\vec{j}_v}\frac{1}{D_v(\vec{j}_v) \left( D_v(\vec{j}_v)+1\right)}.
\end{equation}
Thanks to the factorization 
\begin{equation}
\mathcal{S}^{[\vec{j}_v]}_v=\mathcal{S}^{0[\vec{j}_v]}_v\prod_{i=1}^d\mathcal{S}^{i[j_v^i]}_v
\end{equation} the trace in \eq{big}, for which we use the short notation $\Tr [\cdot]$, splits into the product of three terms involving, separately, bulk ($\dot{\gamma}$), internal link ($L$) and boundary ($\partial \gamma$) degrees of freedom:
\begin{equation}\begin{split}
\Tr [\cdot]=&\Tr_{\dot{\gamma}} \left[\left(\bigotimes_{v: \sigma_v=-1}\bigoplus_{\vec{j}_v} \mathcal{S}^{0[\vec{j}_v]}_v\right)\bigoplus_{J}\mathcal{S}^{[J]}_{\Omega}/\mathbb{I}\right]\cdot 
\Tr_{{L}} \left[ \left(\bigoplus_{J_L}\left(\rho^{J_L}_{L}\right)^{\otimes 2}\right)\bigotimes_{e_v^i\in L: \sigma_v=-1}\bigoplus_{j_v^i} \mathcal{S}^{i[j_v^i]}_v\right]
\\&\cdot \Tr_{\partial\gamma}\left[\left(\bigotimes_{e_v^i\in \partial\gamma: \sigma_v=-1}\bigoplus_{j_v^i} \mathcal{S}^{i[j_v^i]}_v\right)\bigoplus_{J_\partial}\mathcal{S}^{[J_\partial]}_{A}/\mathbb{I}\right]\\=&\Tr_{\dot{\gamma}} \left[\bigoplus_{J}\left(\bigotimes_{v: \sigma_v=-1} \mathcal{S}^{0[\vec{j}_v]}_v\right)~\mathcal{S}^{[J]}_{\Omega}/\mathbb{I}\right]\cdot 
\Tr_{{L}} \left[ \bigoplus_{J_L}\left(\rho^{J_L}_{L}\right)^{\otimes 2}\bigotimes_{e_v^i\in L: \sigma_v=-1} \mathcal{S}^{i[j_v^i]}_v\right]\\&
\cdot \Tr_{\partial\gamma}\left[\bigoplus_{J_\partial}\left(\bigotimes_{e_v^i\in \partial\gamma: \sigma_v=-1} \mathcal{S}^{i[j_v^i]}_v\right)~\mathcal{S}^{[J_\partial]}_{A}/\mathbb{I}\right]
\end{split}
\end{equation}
By introducing a cut-off $j_\text{max}$, so that the trace over a direct sum of matrices can be turned into a sum of traces, we obtain
\begin{equation}\begin{split}
\Tr [\cdot]&\approx\left(\sum^{j_\text{max}}_{J}\Tr \left[\left(\bigotimes_{v: \sigma_v=-1} \mathcal{S}^{0[\vec{j}_v]}_v\right)~\mathcal{S}^{[J]}_{\Omega}/\mathbb{I}\right]\right)\cdot 
\left(\sum^{j_\text{max}}_{J_L}\Tr \left[ \left(\rho^{J_L}_{L}\right)^{\otimes 2}\bigotimes_{e_v^i\in L: \sigma_v=-1} \mathcal{S}^{i[j_v^i]}_v\right]\right)
\\&\cdot\left( \sum^{j_\text{max}}_{J_\partial} \Tr\left[\left(\bigotimes_{e_v^i\in \partial\gamma: \sigma_v=-1} \mathcal{S}^{i[j_v^i]}_v\right)\mathcal{S}^{[J_\partial]}_{A}/\mathbb{I}\right]\right).
\end{split}
\end{equation}
We then write the quantities $\overline{Z_{1/0}}$ as
\begin{equation}\begin{split}
\overline{Z_{1/0}}= \sum_{\vec{\sigma}} \mathcal{C} ~\Tr [\cdot]= \sum_{\vec{\sigma}}  e^{-\mathcal{A}_{1/0}(\vec{\sigma})}
\end{split}
\end{equation}
with
\begin{equation}\begin{split}\label{ageneral}
\mathcal{A}_{1/0}(\vec{\sigma})\coloneqq& -\log \mathcal{C} -\log \left(\sum_{J}\Tr \left[\left(\bigotimes_{v: \sigma_v=-1} \mathcal{S}^{0[\vec{j}]}_v\right)~\mathcal{S}^{[J]}_{\Omega}/\mathbb{I}\right]\right)-\log 
\left(\sum_{J_L}\Tr \left[ \left(\rho^{J_L}_{L}\right)^{\otimes 2}\bigotimes_{e_v^i\in L: \sigma_v=-1} \mathcal{S}^{i[j_v^i]}_v\right]\right)
\\&-\log \left( \sum_{J_\partial} \Tr\left[\left(\bigotimes_{e_v^i\in \partial\gamma: \sigma_v=-1} \mathcal{S}^{i[j_v^i]}_v\right)\mathcal{S}^{[J_\partial]}_{A}/\mathbb{I}\right]\right)
\end{split}
\end{equation}
By introducing pinning fields for boundary ($\vec{\mu}$) and bulk ($\vec{\nu}$) degrees of freedom and performing the traces in \eq{ageneral} we obtain the Ising-like action
\begin{equation}\begin{split}\label{amunu}
\mathcal{A}[\vec{\mu},\vec{\nu}](\vec{\sigma})=& -\log \mathcal{C} -\log \left(\sum_{J}\prod_{v}D_{\vec{j}_{v}}^{\frac{1}{2}(3 + \sigma_v\nu_v)}\right)-\log 
\left(\sum_{J_L}\prod_{e_{vw}^i\in L}d_{j_{vw}^i}^{\frac{1}{2}(\sigma_v\sigma_w -1)}\right)
-\log \left( \sum_{J_\partial} \prod_{e_v^i\in \partial \gamma}d_{j_v^i}^{\frac{1}{2}(3 + \sigma_v\mu_v)}\right)\\=& -\log \mathcal{C}  -\log \left(\prod_{v}\sum_{\vec{j}_v}D_{\vec{j}_{v}}^{\frac{1}{2}(3 + \sigma_v\nu_v)}\right)-\log 
\left(\prod_{e_{vw}^i\in L}\sum_{j_{vw}^i}~d_{j_{vw}^i}^{\frac{1}{2}(\sigma_v\sigma_w -1)}\right)
-\log \left( \prod_{e_v^i\in \partial \gamma}\sum_{j_v^i}~ d_{j_v^i}^{\frac{1}{2}(3+\sigma_v\mu_v)}\right)\\=&-\log \mathcal{C}  -\sum_v\log \left(\sum_{\vec{j}_v}D_{\vec{j}_{v}}^{\frac{1}{2}(3 + \sigma_v\nu_v)}\right)-\sum_{e_{vw}^i\in L}\log 
\left(\sum_{j_{vw}^i}~d_{j_{vw}^i}^{\frac{1}{2}(\sigma_v\sigma_w -1)}\right)
-\sum_{e_v^i\in \partial \gamma}\log \left( \sum_{j_v^i}~ d_{j_v^i}^{\frac{1}{2}(3+\sigma_v\mu_v)}\right)
\end{split}
\end{equation}
where $\mathcal{A}_{1}(\vec{\sigma})$ is given by \eq{amunu} with $\mu_v=-1/+1$ for $v\in /\notin A$, and $\nu_v=-1/+1$ for $v\in /\notin \Omega$, while $\mathcal{A}_{0}(\vec{\sigma})$ by \eq{amunu} with $\mu_v=\nu_v=+1$ $\forall v$. Specifically, $\mathcal{A}_0(\vec{\sigma})$ takes the form 
\begin{equation}\begin{split}\label{a00}
\mathcal{A}_0(\vec{\sigma})=-\log \mathcal{C}  -\sum_v\log \left(\sum_{\vec{j}_v}D_{\vec{j}_{v}}^{\frac{1}{2}(3 + \sigma_v)}\right)-\sum_{e_{vw}^i\in L}\log 
\left(\sum_{j_{vw}^i}~d_{j_{vw}^i}^{\frac{1}{2}(\sigma_v\sigma_w -1)}\right)
-\sum_{e_v^i\in \partial \gamma}\log \left( \sum_{j_v^i}~ d_{j_v^i}^{\frac{1}{2}(3+\sigma_v)}\right),
\end{split}
\end{equation}
and it is minimized by the all-up configuration:
\begin{equation}\begin{split}
\mathcal{A}_{0\uparrow}\coloneqq \mathcal{A}_0(\sigma_v=+1~\forall v)=& -\log \mathcal{C}  -\sum_v\log \left(\sum_{\vec{j}_v}D_{\vec{j}_{v}}^{2}\right)-\sum_{e_{vw}^i\in L}\log 
\left(\sum_{j_{vw}^i}~1\right)
-\sum_{e_v^i\in \partial \gamma}\log \left( \sum_{j_v^i}~ d_{j_v^i}^2\right)\\=& -\log \mathcal{C}  -\sum_v\log \left(\sum_{\vec{j}_v}D_{\vec{j}_{v}}^{2}\right)-\sum_{e_{vw}^i\in L}\log 
\left(j_\text{max}+1\right)
-\sum_{e_v^i\in \partial \gamma}\log \left( \sum_{j_v^i}~ d_{j_v^i}^2\right)
\end{split}
\end{equation}
The action $\mathcal{A}_1(\vec{\sigma})$ instead depends on the particular region $A \cup \Omega$ we are considering. To compute the entropy of the bulk (case $A=\emptyset$ and $\Omega=\dot{\gamma}$) we have to set $\nu_v=-\mu_v=-1$ $\forall v$, and $\mathcal{A}_1(\vec{\sigma})$ takes the form
\begin{equation}\begin{split}\label{a1}
\mathcal{A}_1(\vec{\sigma})=-\log \mathcal{C}  -\sum_v\log \left(\sum_{\vec{j}_v}D_{\vec{j}_{v}}^{\frac{1}{2}(3 - \sigma_v)}\right)-\sum_{e_{vw}^i\in L}\log 
\left(\sum_{j_{vw}^i}~d_{j_{vw}^i}^{\frac{1}{2}(\sigma_v\sigma_w -1)}\right)
-\sum_{e_v^i\in \partial \gamma}\log \left( \sum_{j_v^i}~ d_{j_v^i}^{\frac{1}{2}(3+\sigma_v)}\right)
\end{split}
\end{equation}
Which spin-configuration minimizes it then depends on the combinatorial structure of the graph; in the case $|\partial \gamma|\approx N$ (almost every vertex has a boundary edge) the last sum in \eq{a1} dominates, therefore $\mathcal{A}_1(\vec{\sigma})$ is minimized by the all-up configuration: 
\begin{equation}\begin{split}
\mathcal{A}_{1\uparrow}\coloneqq \mathcal{A}_1(\sigma_v=+1~\forall v)= -\log \mathcal{C}  -\sum_v\log \left(\sum_{\vec{j}_v}D_{\vec{j}_{v}}\right)-\sum_{e_{vw}^i\in L}\log 
\left(j_\text{max}+1\right)
-\sum_{e_v^i\in \partial \gamma}\log \left( \sum_{j_v^i}~ d_{j_v^i}^2\right)
\end{split}
\end{equation}
The bulk entropy is thus given by 
\begin{equation}\begin{split}
\overline{S_2(\rho_{\dot{\gamma}})}\approx \mathcal{A}_{1\uparrow} - \mathcal{A}_{0\uparrow}=\sum_v\log \frac{\left(\sum_{\vec{j}_v}D^2_{\vec{j}_{v}}\right)}{\left(\sum_{\vec{j}_v}D_{\vec{j}_{v}}\right)}
\end{split}
\end{equation}
In the case $|\partial \gamma|\ll N$ is instead the first sum to dominate, thus $\mathcal{A}_1(\vec{\sigma})$ is minimized by the all-down configuration:
\begin{equation}\begin{split}
\mathcal{A}_{1\downarrow}\coloneqq \mathcal{A}_1(\sigma_v=-1~\forall v)= -\log \mathcal{C}  -\sum_v\log \left(\sum_{\vec{j}_v}D_{\vec{j}_{v}}^2\right)-\sum_{e_{vw}^i\in L}\log 
\left(j_\text{max}+1\right)
-\sum_{e_v^i\in \partial \gamma}\log \left( \sum_{j_v^i}~ d_{j_v^i}\right)
\end{split}
\end{equation}
and the entropy is given by 
\begin{equation}\begin{split}
\overline{S_2(\rho_{\dot{\gamma}})}\approx \mathcal{A}_{1\downarrow} - \mathcal{A}_{0\uparrow}=\sum_{e_v^i\in \partial \gamma}\log \frac{\left( \sum_{j_v^i}~ d^2_{j_v^i}\right)}{\left( \sum_{j_v^i}~ d_{j_v^i}\right)}
\end{split}
\end{equation}
Note finally that, since
\begin{align}
&\sum_v\log \left(\sum_{\vec{j}_v}D_{\vec{j}_{v}}^{\frac{1}{2}(3 + \sigma_v\nu_v)}\right)=\frac{1}{2}\sum_{v}\left[(1-\sigma_v\nu_v)\log 
\left(\sum_{\vec{j}_{v}}~D_{\vec{j}_{v}}\right)+(\sigma_v\nu_v +1)\log 
\left(\sum_{\vec{j}_{v}}~D^2_{\vec{j}_{v}}\right)\right],\\
&\sum_{e_{vw}^i\in L}\log 
\left(\sum_{j_{vw}^i}~d_{j_{vw}^i}^{\frac{1}{2}(\sigma_v\sigma_w -1)}\right)=\frac{1}{2}\sum_{e_{vw}^i\in L}\left[(1-\sigma_v\sigma_w)\log 
\left(\sum_{j_{vw}^i}~\frac{1}{d_{j_{vw}^i}}\right)+(\sigma_v\sigma_w +1)\log \left(j_\text{max}+1\right)\right],\\&
\sum_{e_v^i\in \partial \gamma}\log \left( \sum_{j_v^i}~ d_{j_v^i}^{\frac{1}{2}(3+\sigma_v\mu_v)}\right)=\frac{1}{2}\sum_{e_v^i\in \partial \gamma}\left[(1-\sigma_v\mu_v)\log 
\left(\sum_{j_v^i}~d_{j_v^i}\right)+(\sigma_v\mu_v +1)\log 
\left(\sum_{j_v^i}~d^2_{j_v^i}\right)\right],
\end{align}
the Ising action can also be expressed as follows:
\begin{equation}\begin{split}\label{aa}
\mathcal{A}(\vec{\sigma})=& -\log \mathcal{C} -\frac{1}{2}\sum_{v}\left[(1-\sigma_v\nu_v)\log 
\left(\sum_{\vec{j}_{v}}~D_{\vec{j}_{v}}\right)+(\sigma_v\nu_v +1)\log 
\left(\sum_{\vec{j}_{v}}~D^2_{\vec{j}_{v}}\right)\right]\\&-\frac{1}{2}\sum_{e_{vw}^i\in L}\left[(1-\sigma_v\sigma_w)\log 
\left(\sum_{j_{vw}^i}~\frac{1}{d_{j_{vw}^i}}\right)+(\sigma_v\sigma_w +1)\log \left(j_\text{max}+1\right)\right]\\&-\frac{1}{2}\sum_{e_v^i\in \partial \gamma}\left[(1-\sigma_v\mu_v)\log 
\left(\sum_{j_v^i}~d_{j_v^i}\right)+(\sigma_v\mu_v +1)\log 
\left(\sum_{j_v^i}~d^2_{j_v^i}\right)\right].
\end{split}
\end{equation}

\bibliographystyle{jhep}
\bibliography{bibliomaps}
\end{document}